\documentclass[preprint]{aastex}
 
\usepackage{lscape}
\usepackage{natbib}

\begin{document}

\title{Probing the Halo From the Solar Vicinity to the Outer Galaxy: \\
Connecting Stars in Local Velocity Structures to Large-Scale Clouds}

\author{Kathryn V. Johnston\altaffilmark{1,2}, Allyson A. Sheffield\altaffilmark{1}, Steven R. Majewski\altaffilmark{3} and  Sanjib Sharma\altaffilmark{4}}

\altaffiltext{1}{Department of Astronomy, Columbia University, New York 10027, USA}
\altaffiltext{2}{kvj@astro.columbia.edu}
\altaffiltext{3}{Dept. of Astronomy, University of Virginia, Charlottesville, 
VA 22904-0818}
\altaffiltext{4}{Sydney Institute for Astronomy, School of Physics, University of Sydney, NSW 2006, Australia}

\begin{abstract}
Large-scale surveys of the stellar halo have revealed a wealth of substructures, often detected as moving groups relatively nearby the Sun (within a few kpc) or giant spatial inhomogeneities towards the outskirts of the Galaxy (from tens to 100 kpc).
This paper presents the first connections made between two local features in velocity-space and spatial structures on global scales.
The nearby features were discovered in a medium-resolution survey of  M giant stars within $\sim$10kpc of the Sun, apparent as sequences of stars whose radial velocity varied linearly with Galactic longitude.
Comparison to cosmological, chemodynamical stellar halo models confirm that the M giant population is particularly sensitive to rare, recent and massive accretion events, which can give rise to the observed velocity sequences.
The sequences are a signature of debris from a common progenitor, passing at high velocity through the survey volume, near the pericenters of their eccentric orbits.
In the models, the observed stars represent only a small fraction of the original object: the majority of the debris is instead in large structures, whose morphologies are more cloud-like than stream-like and which lie at the orbital apocenters.
 Adopting this interpretation, the full-space motion of the observed stars are derived under the assumption that the members within each sequence share a common velocity.
Orbit integrations are then used to trace the past and future trajectories of these stars across the sky, in distance and in line-of-sight velocity.
The predicted paths reveal plausible associations with large, cloud-like structures previously discovered as spatial inhomogeneities in the Sloan Digital Sky Survey \citep[Hercules-Aquila Cloud ---][]{belokurov07} and the Two Micron All Sky Survey \citep[structures A11 and A13 recovered by][]{sharma10}.
While stellar {\it streams} around our Galaxy have been extensively mapped and modeled, a corresponding understanding of these (and other) debris {\it clouds} is much less well-developed.
The connections made between nearby velocity structures and these distant clouds represent preliminary steps towards developing more coherent maps of such debris systems.
These maps promise to provide new insights into the origin of debris clouds, new probes of Galactic history and structure, and new constraints on the high-velocity tails of the local dark matter distribution that are essential for interpreting direct detection experiments.
\end{abstract}


\clearpage

\section{Introduction}
\label{intro.sec}
There is a long history of using individual stars 
to map Galactic structure.
Studies have gradually evolved from understanding the gross structure of the Galaxy and our place within it
\citep[e.g.][]{herschel,kapteyn22,shapley28},
to clarifying the number and character of major structural components --- for example 
debating the existence of distinct thin and thick disk components \citep[see, e.g.,][] {gilmore89,majewski93} and the nature and shape of the Galactic bulge \citep[e.g.,][]{weinberg92}, with its potential x-like structure \citep{mcwilliam10,saito11} and perhaps multiple bars \citep{nishiyama05}.
The last two decades have seen vast increases in the sizes of stellar photometric catalogues which have allowed this field to grow to include the study of {\it substructure} within the Galaxy.
Stars selected from large scale photometric surveys --- such as the Sloan Digital Sky Survery \citep[SDSS;][]{newberg02,belokurov06a} and the Two Micron All Sky Survey \citep[2MASS;][]{majewski03,rochapinto03,rochapinto04,rochapinto06} ---  have revealed a wealth of spatially coherent overdensities in the stellar halo beyond the disk of our Galaxy. 
The substructure is dominanted by the tidal streams from the ongoing destruction of the Sagittarius dwarf galaxy, which arc dramatically over the Galactic poles \citep{ibata01,majewski03,belokurov06a}. 
There are also numerous smaller streams from both long-dead satellites \citep[e.g. the Orphan Stream ---][]{belokurov06a} and globular clusters  \citep[e.g.][]{odenkirchen03,grillmair06}.
Less-well studied, but also ubiquitous in these photometric surveys are amorphous clouds of debris --- the most prominent being the Triangulum-Andromeda and the Hercules-Aquila Clouds \citep[TriAnd and HerAq, see][]{majewski04,rochapinto04,belokurov07} and the Virgo and Pisces Overdensities \citep[VOD and POD, see][]{juric08,sharma10}.

Substructure has also been found in other dimensions --- indeed
VOD and POD were seen as groupings in velocity space along limited lines of sight \citep{newberg02,watkins09,kollmeier09} before their full extent was subsequently revealed with photometric maps.
The discovery of similar {\it moving groups} in the disk \citep[e.g.][]{eggen77} predates the modern studies \citep[see summary in][]{majewski94}, and the origin of these disk substructures has been interpreted as either relic associations from their common birth places in ancestral stars clusters or due to stars trapped in resonances with disk and spiral arms \citep{dehnen00,bovy10}.
In the halo, moving groups are thought to be the signature of the disruption of accreted objects. 
Such associations have been found locally in full space motions \citep[e.g.,][]{majewski94,majewski96,helmi99,kepley07,morrison09} and on larger scales using line-of-sight velocities alone \citep[e.g.,][]{ibata95,newberg02,klement09,schlaufman09,williams11}.
In many cases, the significance of tentative associations has been bolstered by similarities in the chemical abundances of member stars \citep[e.g.,][]{majewski96,chou07,chou11}, or clear stellar-population sequences in color-magnitude diagrams \citep[e.g.,][]{williams11}.

This plethora of discoveries has inspired concurrent theoretical work to combine our understanding of debris dynamics, hierarchical structure formation and the evolution of stellar populations and use the observations to: 
reconstruct the history of individual objects \citep[e.g., Sgr --- see][]{johnston95,velasquez95,johnston99b,helmi01,law05,law10}; 
 confirm our expectations for the level of substructure from hierarchical models of structure formation \citep{bullock01,bullock05,cooper10a,rashkov11} in space \citep{bell08,sharma10}, velocities \citep{harding01,xue10,helmi11} and stellar populations \citep{font06,font08,bell10};
measure the depth, radial profile and shape of the Galactic potential \citep[e.g.,][]{johnston99a,ibata01,helmi04,johnston05,law09,koposov10};
and  search for the presence of dark matter substructure \citep[i.e., the ``missing satellites'' of][]{moore99,klypin99} within the Galactic halo \citep{ibata02,johnston02,yoon11}.
\footnote{Note: analogous work has also been done on studies of resolved stellar populations around M31 \citep[see][]{ferguson02,ibata07,gilbert09}.}

This paper builds on the understanding of stellar halo substructure developed over the last two decades, bringing together the underlying themes of discovery, connection and interpretation.
This work is inspired by the combined implications of three distinct studies.
The first study explored the morphologies and origin of substructure in a model of a stellar halo that had been built by superposing simulations of individual accretion events \citep{bullock05}. 
Debris structures with both stream-like and cloud-like morphologies were apparent. In some cases, there were two or even three distinct clouds associated with a single disruption event  --- the analogues of ``shells'' seen around other galaxies \citep{malin83,quinn84} but viewed from an internal rather than external perspective.
The clouds represented debris slowly turning around at the apocenters of orbits that were typically  more eccentric orbits than those of the debris  streams \citep{johnston08}.
The implication of this work is that there must be low-density stellar streams moving between these {\it apocentric clouds} and passing through the inner Galaxy at high speed and approaching the pericenters of these eccentric orbits --- we will hereafter refer to streams of this type as a {\it pericentric streams}  to distinguish them from tidal streams that are apparent in spatial maps at much higher density.

In the second study \citep[][hereafter Paper I]{sharma10}, a group finder was  applied to the 2MASS M giant catalogue on large scales ($10< K_s$, corresponding to distances $d$ in the range $\sim 10-100$ kpc ). The group finder recovered many of the already known substructures, including Sagittarius, provided the first spatial map of POD that revealed its cloud-like morphology, and identified several other amorphous overdensities, whose nature has yet to be confirmed. 
 
The third study  \citep[][hereafter Paper II]{sheffield12} was a survey of 1799 line-of-sight velocities (RV's)  of relatively nearby ($K_s<9$, with distances within $\sim$10 kpc), moderate Galactic latitude ($30^\circ < |b| < 60^\circ$) M giant stars selected from the 2MASS catalogue.
As anticipated, the survey was dominated by thick disk stars, but also showed suggestive velocity sequences in the smaller populations of ``RV-outliers'' (i.e., those with high velocities relative to the Sun).
Follow-up high-resolution spectroscopy of samples of these RV outliers revealed that many had abundance patterns consistent with an accreted population and inconsistent with one formed early on in the Galactic halo or disk, and therefore lending some support to the proposal of a real physical association between stars in the suggestive sequences.

In this paper, we look at an intriguing question raised by these studies: whether the sequences of  M giant RV-outliers identified in the inner Galaxy in Paper II actually represent {\it pericentric streams} between the {\it apocentric clouds} of M giants identified in the outer Galaxy by the group finder in Paper I.
While neither the inner-Galactic velocity sequences nor the outer-Galactic spatial overdensities may be entirely convincing as real physical associations in themselves, exploring possible connections between the two at least allows us to make predictions of observable properties these tentative structures must fulfill if they {\it are} related, and these predictions can be tested with future surveys. 

Any successful attempt at making these connections --- from pericentric streams to apocentric clouds or even from clouds-to-clouds --- could have far broader implications.
In both observations and models of stellar halos, debris clouds occur with a similar frequency to streams, yet the former have received far less attention.
These connections would provide the first comprehensive maps of systems creating cloud-like debris. 
Indeed, they would be among the first maps of {\it any} debris system to explore the full range of orbital phases, from pericenter to apocenter:
of all the substructures around our Galaxy, only the streams from Sgr have been followed for more than one radial orbit.
The great extent of Sgr's debris has allowed a particularly detailed view of its history, as well as constraints on the depth and triaxiality of the Milky Way's potential \citep[e.g.]{johnston99b,ibata01,helmi04,johnston05,law09,law10}.
Extensive maps of debris clouds could provide similar constraints on their individual histories,  on the Galaxy's accretion history and on the Galaxy's mass distribution.
Moreover, conclusive evidence for and measurements of the velocity of streams of stellar debris passing at high speed through the solar vicinity between these clouds could be used to help predict and interpret signatures of the associated dark matter in direct detection experiments, contributing to what we can say about the nature of dark matter \citep[see e.g.][]{kuhlen12}.

In Section \ref{data.sec} we review the observational and model data utilized in the paper. 
In Section \ref{sims.sec} we ``observe'' our model stellar halos to understand the strengths and limitations of our observational survey, and develop and test our interpretation methods.
In Section \ref{obs.sec} we apply the methods developed in Section \ref{sims.sec} to the stellar sequences in our survey and compare our predictions to observed data sets.
We summarize our conclusion in Section \ref{conclusions.sec}

\section{Data Sets}
\label{data.sec}

\subsection{The M Giant Survey}
\label{mgiant.sec}


The M giant survey described in Paper II has two goals: (1) to study the bulk characteristics of the thick disk and (2) to study the nearby halo, in particular the accreted component.
Following \citet{majewski03}, M giant candidates were selected from the Two Micron All Sky Survey  on the basis of their color: 96\% have $J-K_s > 0.85$.
To isolate mainly thick disk and nearby halo stars, cuts in Galactic latitude of $30^{\circ}<|b|<60^{\circ}$ and apparent magnitude were imposed.
The range in $K_{S,0}$ spans 4.3$<K_{S,0}<$12.0, with a mean of 7.5$\pm$1.3.
Spectra of the candidates were taken using the medium-resolution FOBOS spectrograph on the 1-m telescope at the Fan Mountain Observatory and the Cassegrain spectrograph on the 1.5-m telescope at Cerro Tololo Inter-American Observatory ---
 the results for the first  1799 stars are presented in Paper II.
The radial velocities have an estimated error of 5-10 km s$^{-1}$, based on repeat observations and also on comparisons with followup high-resolution spectra.

\begin{figure}
\epsscale{0.9}
\plotone{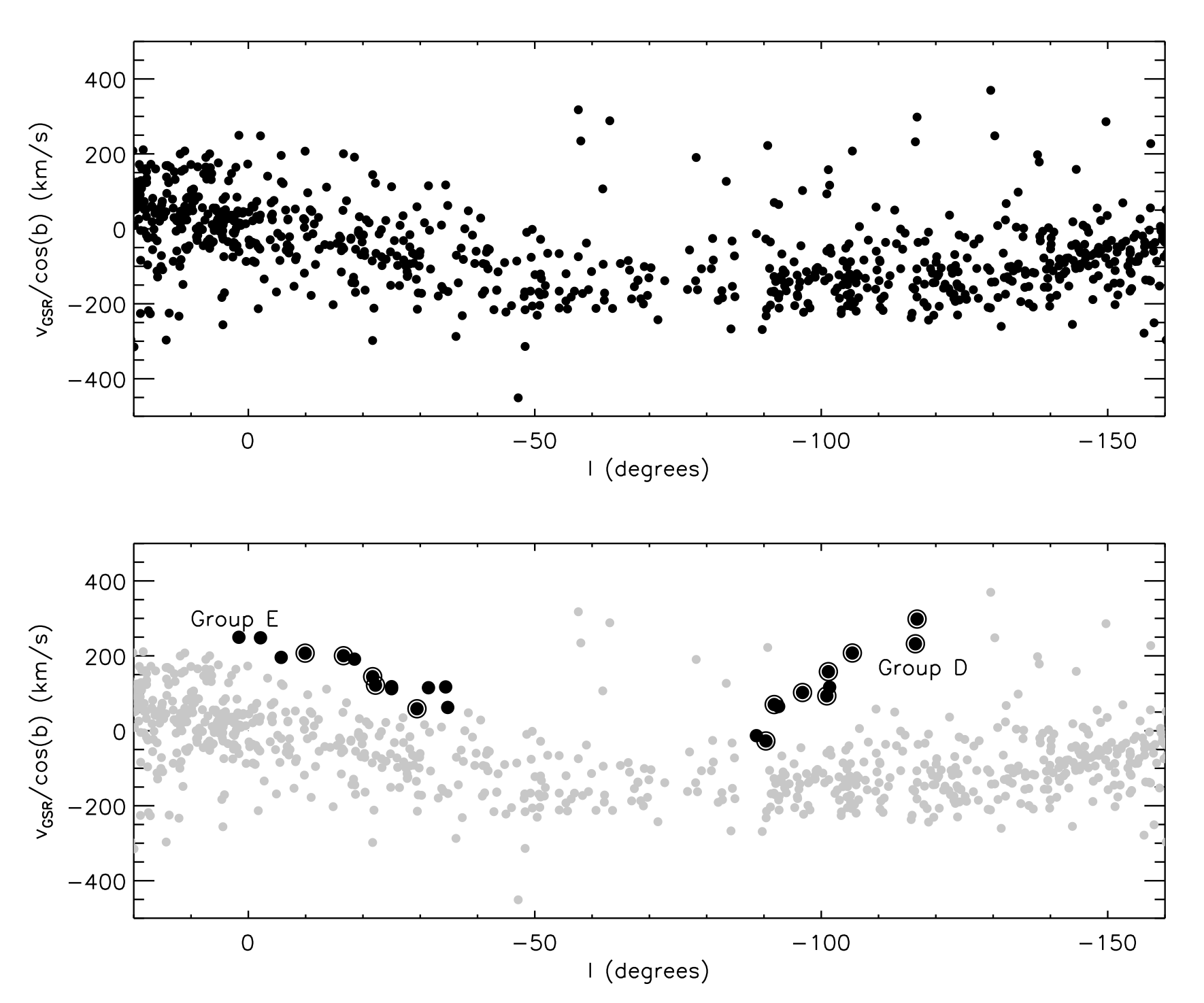}
\caption{The radial velocity distribution of our M giant survey. Both panels show the line-of-sight velocities projected to the disk plane by a $\cos b$ normalization, in the Galactic longitude range $-20^\circ < l < 160^\circ$. The black symbols in the lower panel correspond to stars selected to be part of the two groups chosen for detailed study here. Those points enclosed in circles represent stars having high-resolution spectra and abundance measurements. 
\label{obs.fig}
}
\end{figure}

Figure \ref{obs.fig} summarizes the data by plotting the line-of-sight velocity (RV) in the Galactic Standard of Rest ($v_{\rm GSR}$) frame as a function of Galactic longitude $l$. The velocities are divided by $\cos(b)$ (where $b$ is Galactic latitude) in order to project disk stars moving at similar velocities in the disk plane into a tighter distribution. Many RV outliers are apparent outside the disk distribution, some making suggestive sequences in this plane.
Follow-up, high resolution spectroscopy of 34 of these RV outliers in Paper II showed that many had abundance patterns consistent with a recently-accreted population with low metallicity and $\alpha$-element abundances reminiscent of surviving Milky Way satellite galaxies. 
The abundance results lend some support to the interpretation of the sequences as the signature of real physical associations and motivate the work in this paper.
Two particular sequences (referred to as Groups D and E in Paper II) that are analyzed in detail in this paper are highlighted in black in  the lower panel of Figure \ref{obs.fig}, with those with high-resolution spectra outlined by circles.

\subsection{The Simulated Survey}
\label{simsurvey.sec}

We compare our M giant data to the 
eleven stellar halo models described in \citet{bullock05}.
The models were built entirely from accretion events drawn from histories representing random realizations of the formation of a Milky-Way-type galaxy in a $\Lambda$CDM Universe.
The phase-space structure in the models was constructed by superposing the final positions and velocities of particles at the end point of individual N-body simulations of dwarf galaxies disrupting around a parent galaxy matching the cosmological model accretion event history.
The particles in each of these simulated objects had equal dark matter masses and were given (varying) associated stellar masses in such a way that the luminous material reproduced the structural scaling relations observed for Local Group dwarfs.        
Star formation histories were assigned to the star-particles within each accreted dwarf using a simple leaky-accreting-box model of star formation and chemical enrichment that was abruptly truncated upon accretion \citep{robertson05,font06}.

We created simulated surveys of our model halos using the {\it
  Galaxia} code \citep{sharma11b}, which can generate a mock 
observational survey from a given N-body model.
The code takes the star formation and chemical evolution histories of
the satellites in the models and uses Padova isochrones 
\citep{bertelli94,giradi00,marigo07,marigo08}, 
to generate the observable properties of these stellar populations. 
A mock M giant survey was created by applying the same cuts to the models as to the 2MASS catalogue --- selecting stars: in the latitude range $30^\circ < |b| < 60^\circ$; with  $(J-K_s) > 0.85$ and $0.22< (J-H) - 0.561(J-K_s) < 0.36$; and apparent magnitudes $K_s < 9$.

We caution that our simulated surveys are not expected to provide accurate {\it predictions}  of the structure of stellar halos in local volumes around the Sun for several reasons: (i)
there is no component of stars formed either {\it in situ} in the main halo or kicked out from the Galactic disk, both of which could be a significant contributor in this region \citep[see][for discussions of these formation mechanisms for halo stars]{abadi06,zolotov09,purcell10,zolotov10,font11}; (ii) 
the parent galaxy was represented by analytical functions in the N-body simulations, so there was no mixing of stars due to violent relaxation during significant mergers; (iii)
the models were constructed before the discovery of the ultra-faint satellites \citep[e.g.]{willman05,belokurov06b}, so only contain contributions from more luminous satellites.
and (iv) due to limited numerical resolution of the 
  N-body models, Galaxia has to oversample the N-body particles when 
  generating stars, the oversampled stars need to be dispersed in phase space 
this degrades the sharpness of velocity  structures (in the generated mock data the ratio of number of
stars to the number of N-body particles was about 2).

Nevertheless, while a direct comparison of (say) the ubiquity or mass spectrum of substructure in the data and the models is not relevant, these models are useful for developing some intuition on the meaning of our observational results. 
In particular, in Section \ref{sims.sec} we use them to illustrate how our selection of stellar populations could change our view of the local halo: e.g. what substructures in our data might represent and what these substructures could be telling us about the formation and global structure of the stellar halo. 
We then use this understanding to go on in Section \ref{obs.sec} to interpret the nature of the observed structures.

\section{Results I: Examination of the Synthetic Surveys}

\label{sims.sec}

\begin{figure}
\plotone{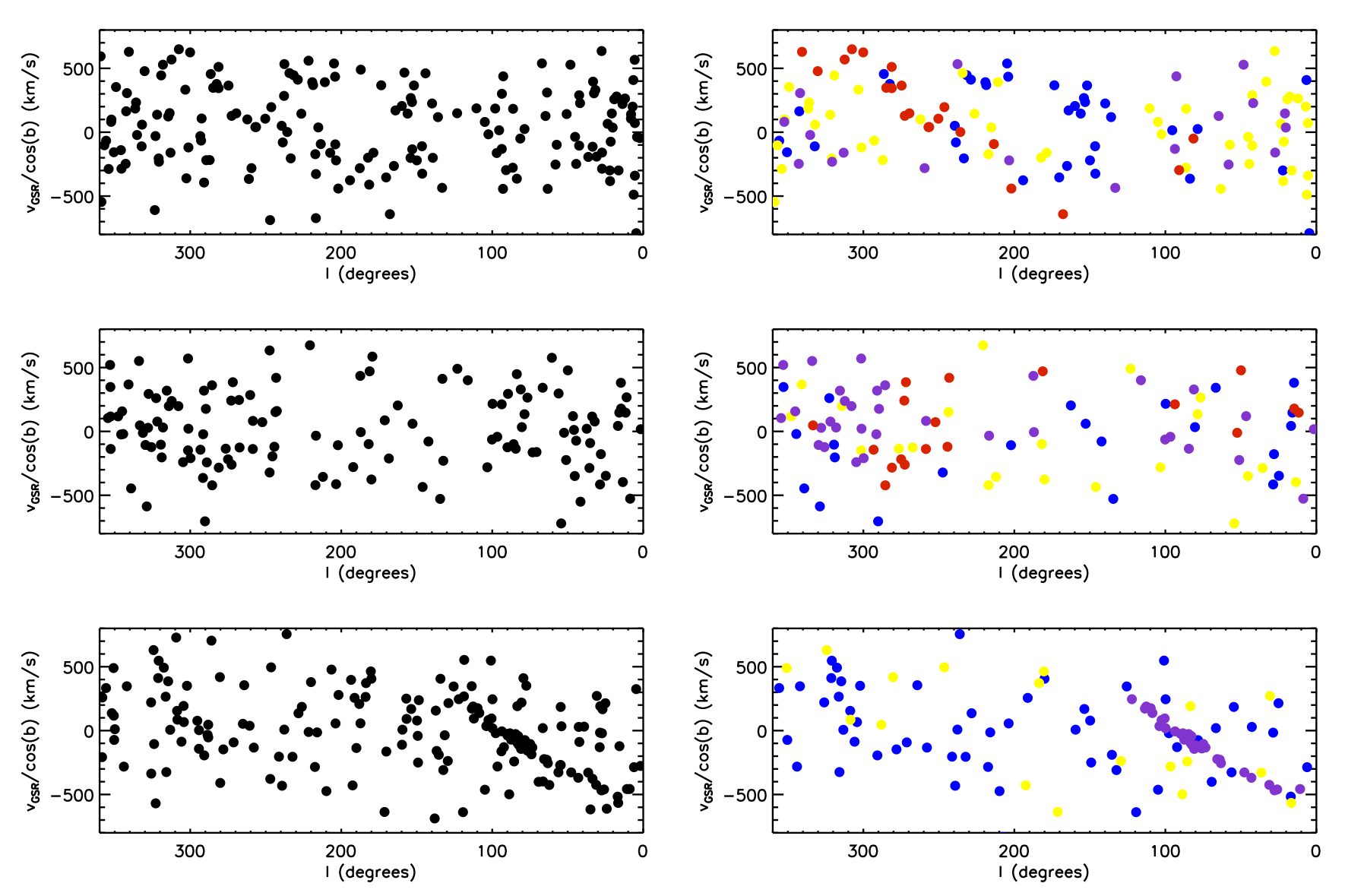}
\caption{As Figure \ref{obs.fig} but for our synthetic surveys M giants for three example simulated halos (left-hand panels). Only 1-in-10 of the synthetic M giants were plotted to show a number of RV outliers comparable to those seen in Figure \ref{obs.fig}.
In the right-hand panels, only particles from satellites contributing more than 10\% of the stars to each synthetic survey are plotted,  color coded by satellite progenitor. Note the prominent velocity sequences features in red in the top panel (``Group 1'') and purple in the bottom panel (``Group 2'').
\label{vrvsl.fig}
}
\end{figure}

The left-hand panel of Figure \ref{vrvsl.fig} shows three examples of our simulated surveys projected onto the same plane as the observations in Figure \ref{obs.fig}. Note that there is no disk component in our synthetic surveys, and only 1-in-10 of our simulated M giants are displayed  to mimic the approximate number of RV-outliers present in the data taken in the real survey to date. All panels illustrates the general result of these surveys: stars are mostly distributed randomly in the plane, but with some distinct groups and sequences across spans of tens of degrees in Galactic longitude $l$. 
In particular, the upper and lower panels contain examples of distinct linear sequences (in red and purple respectively, hereafter referred to as Groups 1 and 2), reminiscent of the structure of the some of the groups selected from the observations. Further examination of the models showed that 72 \% of the stars in Group 1 and  53\% in Group 2 had [$\alpha$/Fe] $<0.1$ --- clearly offset to low $\alpha$-element abundances compared to the bulk of the model stellar halos from which they were drawn (which both had ~30\%  [$\alpha$/Fe] $<0.1$).
This offset is  in the same sense as the results of the analysis of the high-resolution spectra for the observed RV-outliers in Paper II, which showed that the majority of these stars also had lower $\alpha$-element abundances compared to the standard stellar halo sequence.

\subsection{Biases Inherent in our Group Selections}
\label{sims_bias.sec}

Before looking more closely at the properties of the Groups 1 and 2 in our synthetic surveys it is useful to consider how the nature of the survey itself could bias our view of the stellar halo. 
M giants are relatively rare in the stellar halo because these stars have to be fairly metal rich to evolve into such late type stars. 
This is in part why debris from the Sagittarius dwarf galaxy is so obvious in M giants --- this tracer in particular enhances the contrast of Sgr members against the (generally more metal poor) background \citep{majewski03}.
The top panel in Figure 6 of \citet{sharma11a} provides a graphic illustration of how the M giant populations in our model stellar halos are dominated by luminous events with moderate accretion times.
Only satellites accreted more recently than $\sim$10 Gyrs ago, with luminosities greater than $\sim 10^8 L_\odot$ can each contribute more than 1\%  to the entire M giant  population in a given stellar halo.

Note that the same selections that bias the samples from our models towards moderately recent, more massive accretion events would bias 
the real data set away from the some components that are actually missing from our models.
First, the ultrafaint dwarfs are simply too low luminosity and metallicity to contain significant numbers of M giants.  
In addition, our groups were chosen because they stood out as coherent structures in velocity, distinct from the dominant disk population.
Therefore, by construction these stars must be on eccentric orbits with apocenters well outside the Solar Circle, which imposes a bias against the stars originally formed {\it in situ} in the main halo or in the Galactic disk --- contributions that are also missing from the models.
Indeed, in Paper I we found only six of our 34 RV-outlying stars had properties that indicated they might have been born in the disk, while sixteen clearly resembled our expectations for an accreted population.
Hence, despite their simplicity, our models may in fact provide a fair representation of this particular dataset.

Finally, requiring a sufficient density of M giants in the volume to detect a sequence imposes further constraints:
while stellar streams do get dynamically colder over time , they also get more diffuse \citep[i.e., to conserve overall phase-space density; e.g.,][]{helmi99}, which limits
how ancient an event these detectable groups can come from. 

In summary, we conclude that we expect the groups of halo stars identified in our M giant sample to come pre-dominantly from a few relatively high-luminosity, moderately recent disruptions of satellites along eccentric orbits. Our sample selection biases against stars: (i) formed {\it in situ}; (ii) accreted from ancient events or in low-luminosity progenitors; and (iii) on only mildly-eccentric orbits.

\subsection{Nature of Velocity Structures}
\label{sims_sub.sec}

The expectations from Section \ref{sims_bias.sec} are confirmed with a further dissection of our synthetic samples. The right-hand panels of Figure \ref{vrvsl.fig} repeat the left-hand panels, but with the points color coded according to the satellite from which they came (only satellites contributing more than 10\% of stars in the survey volume are plotted). These panels reinforce the validity of the suggestion in Paper II that many structures picked out by eye from the data could correspond to real physical associations. A wide variety of morphologies of the distribution of stars from a single progenitor in this plane can be seen: note in particular that Groups 1 and 2 indeed correspond to single satellites  (e.g., in red/purple in top/bottom right-hand panels); but also that other groups and sequences with a variety of morphologies also correspond to real associations (e.g., clumps of blue particles in upper panel).

As expected, the satellite progenitors of Groups 1 and 2 were both fairly luminous and accreted relatively recently --- with stellar  masses and accretion times of ($2.3\times 10^8 L_\odot, 4.6 $ Gyrs) and ($1.2\times 10^7L_\odot, 6.5$ Gyrs) respectively. Given the lower luminosity of its progenitor, the high density of Group 2 in the observational plane is particularly striking: it contributes $\sim$20\% of the total number of M giants to its synthetic survey, while Group 1 contributes only $\sim$10\%. This difference can be explained by looking at the satellite disruption histories. The progenitor of Group 1 was entirely disrupted more than 0.63 Gyrs ago, presumably on the previous pericentric passage. In contrast, 70\% of the stars in the progenitor of Group 2 became unbound less than 0.1 Gyrs ago, and most of those stars overlap our survey volume. This suggests that, while Group 1 is likely composed of free-streaming stars largely unaffected by self-gravity, Group 2 is observed during the final stages of disruption on the current pericentric passage and while self-gravity may still be important (i.e. similar to the Sgr dwarf). Overall, we conclude that Group 2 does not provide a good model for our observed groups in our M giant survey, both because of it's density (in the observed data, Groups D and E contain 11 and 13 stars respectively, both less than 5\% of the stars with line-of-sight velocities greater than 100 km/s relative to the Sun) and because our synthetic surveys are catching its progenitor at a special short-lived phase of its evolution. 

\begin{figure}
\plotone{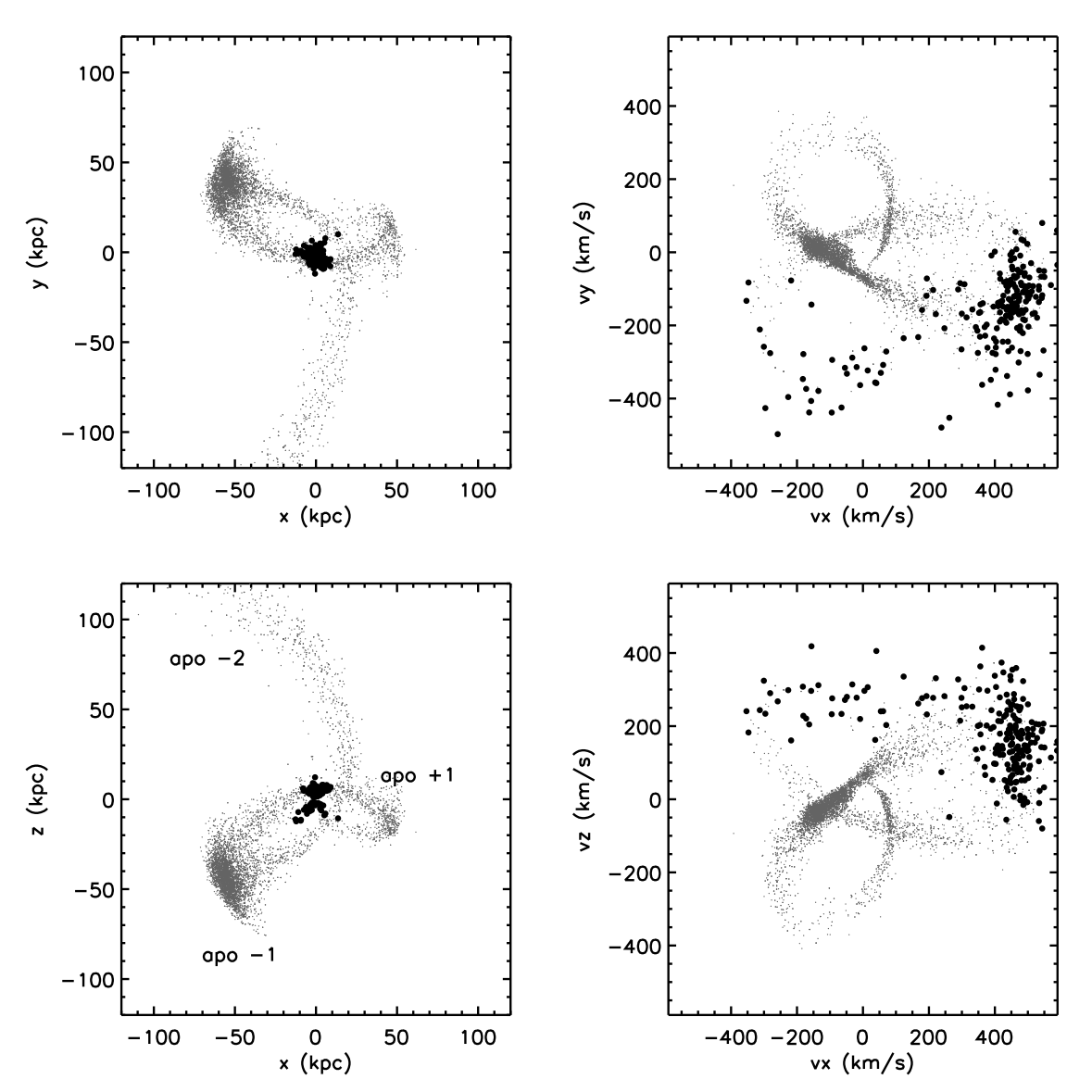}
\caption{Position- (left hand panels) and velocity- (right-hand panels) space projections of entire stellar population associated with Group 1 in the simulations (gray points), with those that fell in the survey volume highlighted in black.
The $(x,y,z)$ directions point from the Sun to the Galactic center, in the direction of Galactic rotation and towards the North Galactic Pole respectively. In Galactocentric co-ordinates, the Sun sits at $(x,y,z)=(-8,0,0)$kpc.
\label{sats.fig}
}
\end{figure}

The remainder of the section concentrates on Group 1 as the closest analogue in our synthetic surveys to the groups observed in the M giant data set.
Figure \ref{sats.fig} explores how the stars in Group 1 are related to the full phase-space distribution of debris for its progenitor satellite . The left/right panels show the position/velocity distribution of all particles in gray, with those that fell in our sample volume highlighted in black. The stars in the survey tend to be sitting near the pericenters of their orbits and traveling at high velocities.  The linear feature arises from a single stream of stars in the vicinity of the Sun moving with a small spread around a single velocity vector. (There are actually two sets of particles present from two different pericentric passages within the survey volume, but one is clearly dominant.)

Perhaps most striking about Figure \ref{sats.fig} is how the debris morphology contrasts with the classical stellar streams that have been detected in photometric surveys \citep[e.g., the view of Sagittarius and the Orphan Stream provided by SDSS][]{belokurov06a} ---
likely a consequence of our selection of RV outliers stars which biases our chosen groups towards debris on eccentric orbits.
The fraction of stars in the pericentric velocity stream is tiny --- it would be too diffuse to be detected in space alone.
Indeed, the majority of the associated debris is not in stream-like structures at all, but is instead lying in large clouds of debris around the apocenters of the orbit: more than 85\% of the stars are at distances greater than 45kpc from the center of the Milky Way. 
These clouds are also very diffuse since they subtend large angles at the Milky Way's center, but, in constrast to the pericenter stream, they are in a very low-density region of the Galaxy.

The following sections explore the intriguing possibility of using the few stars observed in our pericentric streams to search for the rest of the disrupted satellite, in particular in apocentric clouds.

\subsection{Derivation of Full Space Motion from Line-of-sight Trends}
 \label{motion.sec}
Suppose the observed groups indeed corrsepond to pericentric streams of stars moving with a single velocity $\vec{V}=(V_x,V_y,V_z)$ through our heliocentric survey volume, where the $(x,y,z)$ directions are towards the Galactic center, in the direction of Galactic rotation and towards the North Galactic Pole respectively. 
(Note, many observational groups report their results in a co-ordinate system $(U,V,W)=(-v_x,v_y,v_z)$ and we reserve these symbols $(U,V,W)$ to refer to this alternative convention when comparing with their work later in the paper.)
A star in the group at Galactic longitude $l$ and latitude $b$ would have line-of-sight velocity
\begin{equation}
  v_{\rm r, predicted}=V_x \cos(b) \cos(l)+V_y \cos(b)\sin(l) +V_z \sin (b). 
\end{equation}
Some of the sequences in the $(l,v_r/\cos(b))$ plane in our real  and synthetic surveys could represent part of the sinusoid in $l$ derived from this equation, whose phase and amplitude is determined by the values of $V_x$ and $V_y$. For a perfectly cold stream, any deviation from the  mean trend with $l$ would be due to the $V_z \tan (b)$ term --- hence it would be possible to deduce the three unknowns, $\vec{V}$, with RV measurements of just three stars in a group spread over a variety of directions in the sky. This approach of using projected motions from several lines of sight to determine a full-space motion is equivalent to using  perspective rotation to determine the proper motion of extended objects \citep{feast61} which has been applied to nearby objects \citep{merritt97}, M31 \citep{vandermarel08} and even clusters of galaxies \citep{hamden10}.

In reality our ability to recover the full space motion could be compromised by: (i) incomplete sky coverage; (ii) small numbers of stars; (iii) intrinsic stream dispersion; and (iv) the presence of velocity gradients in the stream within our survey volume. 
The last of these effects 
becomes important once the expected change in velocity 
\begin{equation}
	\Delta v \sim{\rm local \;\; acceleration} \times {\rm length \;\; of \;\; stream \;\; in \;\; volume \over stream \;\; velocity}\sim {v_{\rm circ}^2 \over R_\odot }\times {\Psi \; d \over v_{\rm stream}}
\end{equation}	
is significant
given the number of stars and stream dispersion.
Here $v_{\rm circ}$ is the speed of a closed orbit at the Solar circle $R_\odot$ and  $(d, \Psi, v_{\rm stream})$ are the distance, angular extent and speed of stars along the stream respectively.
Our observed data groupings have $n_* < 15$ members, spanning ranges of $< 30^\circ$ on the sky, at distances of 3-8 kpc from the Sun and with line-of-sight speeds of order several hundred km/s. Hence, rescaling parameters, we find that
\begin{equation}
\Delta v	\sim  45 {\rm km/s} \times \left({v_{\rm circ} \over 236 {\rm km/s}}\right)^2
	 \times \left({8 {\rm kpc} \over R_\odot}\right)
	  \times \left( {400 {\rm km/s} \over v_{\rm stream}} \right)
	  \times  \left({\rm \Psi \over 30^\circ}\right)
	  \times \left({\rm d \over 5 {\rm kpc}}\right).
	  \end{equation}
While our measurement errors on individual velocities are typically smaller than $\Delta v \sim 45$km/s, we nevertheless expect our overall random uncertainty (e.g. due to stream dispersion) to be greater than this for our small data sets.
In future work, either local accelerations (i.e. $a_r\sim v_{\rm circ}^2/R_\odot$) could be used to analytically account for velocity gradients across the survey volume, or the velocity parameters could be used as initial conditions for a full orbit integration in some assumed Galactic potential.

\begin{figure}
\epsscale{0.5}
\plotone{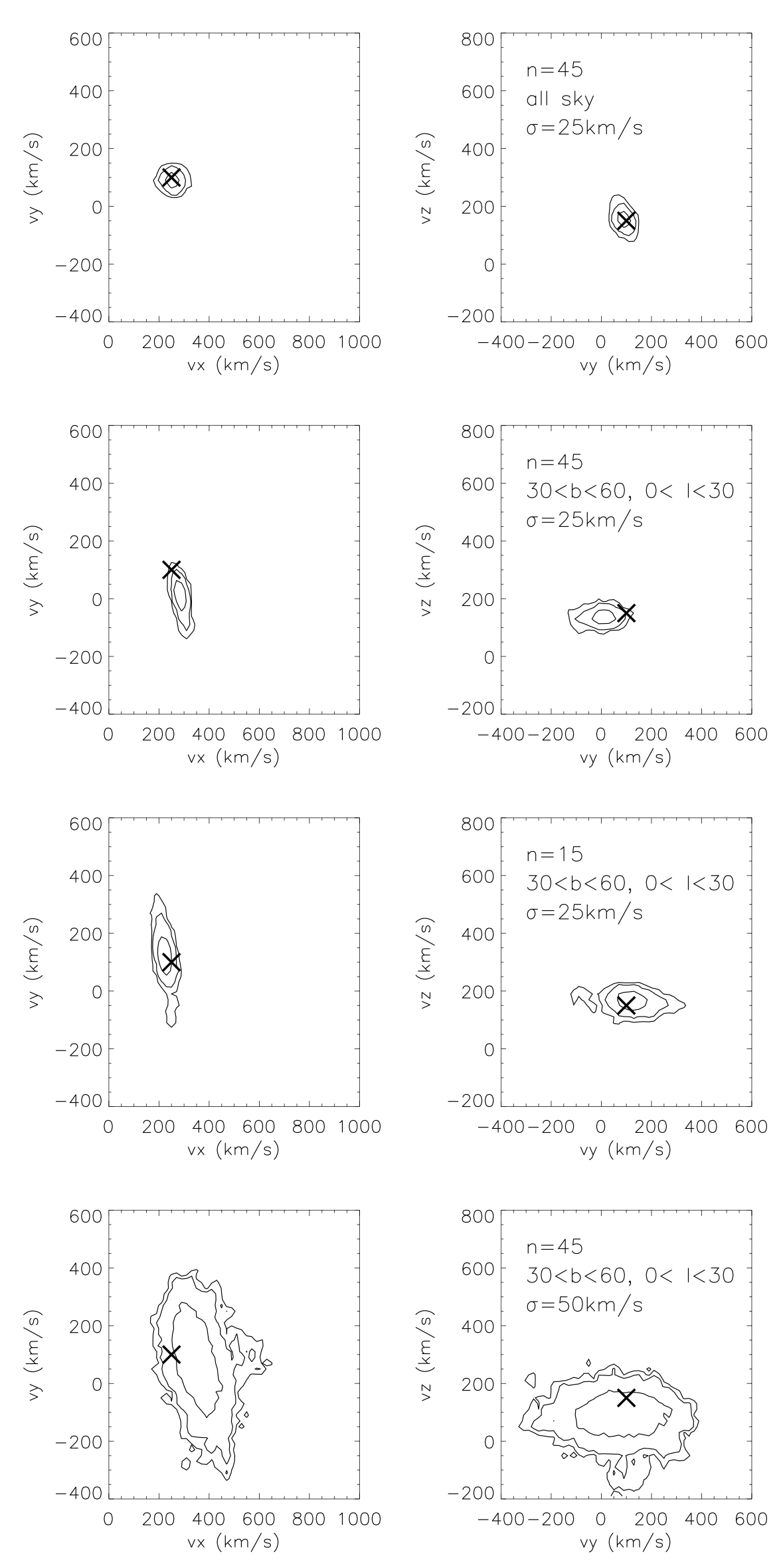}
\caption{Projections of the PDF for recovered full-space motion from applying the MCMC method to our ``ideal'' samples, moving with a single mean velocity (indicated by the cross in each panel) through the survey volume.
The velocities are scattered about the mean by drawing at random from a Gaussian distribution of dispersion $\sigma$ in each direction.
The crosses indicate the true average motion of the sample. The different rows contrast varying sky-coverage, number of stars and dispersion (labeled in the right-hand panels).
\label{ideal_mcmc.fig}
}
\end{figure}

The effects of the remaining three of limitations mentioned above were assessed by applying a Monte Carlo Markov Chain (MCMC) algorithm to synthetic observations of {\it idealized} streams, which were constructed (via random draws from analytic functions) to  conform strictly to our assumptions of no velocity gradient in the survey volume and an isotropic stream dispersion. 
The MCMC algorithm followed the steps outlined in \citet{verde03} assuming a likelihood
\begin{equation}
  L(\vec{V},\sigma)={1\over \sqrt{2 \pi}\sigma} \Sigma_{i=1}^{n_\star}\exp\left(-{(v_{r,i}- v_{\rm r, predicted}(l,b))^2 \over 2 \sigma^2}\right).
\end{equation}
of the parameters (average velocity $\vec{V}$ and isotropic stream dispersion $\sigma$) given the data ($(v_r,l,b)$ for $n_\star$ stars).
The chains typically took $\sim$5000 steps to burn-in and the algorithm was allowed to run a minimum of 20,000 steps to fully explore parameter space.
The chains were terminated when they jointly satisfied the convergence criteria outlined in \citet{verde03}.
Figure \ref{ideal_mcmc.fig} shows projections of the probabality density functions (PDF's) of recovered parameters derived from the distribution of points in the MCMC chains following the burn-in phase. 
Each row represents PDF's of a different set of observations of stars in the {\it idealized} streams whose true motions are indicated by the crosses.
The sets differed in sky coverage, number of stars and intrinsic stream dispersion (as labeled in the right panel in each row).
As might be expected, the contour {\it spacing} increases (and the uncertainty in the parameters goes up) as the number of stars goes down (e.g., comparing the second and third rows) or the dispersion in the stream goes up (comparing the second and fourth row).
The {\it shape} of the contours depends on the sky-coverage of the data points. 

\begin{figure}
\plotone{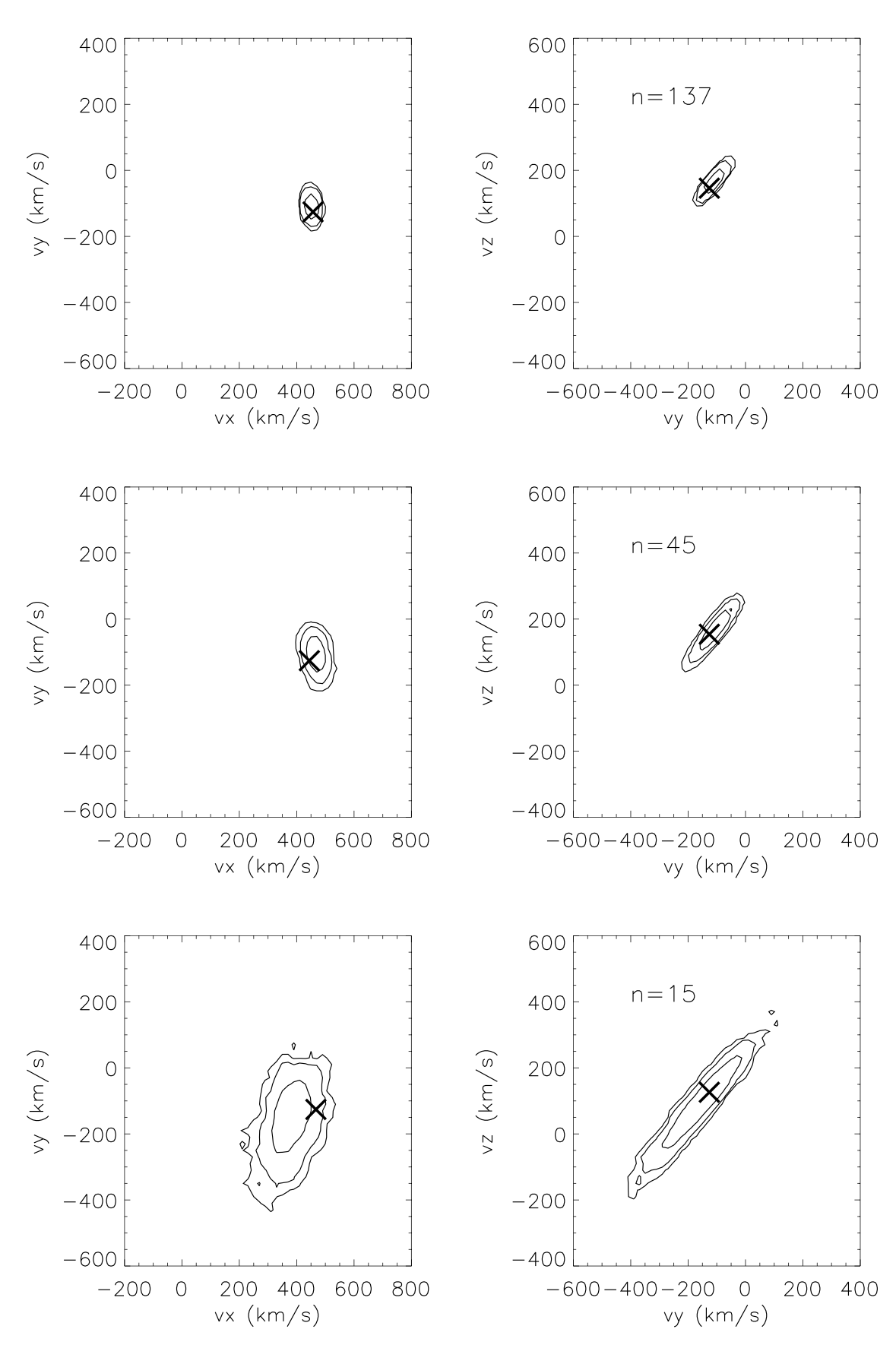}
\caption{As Figure \ref{ideal_mcmc.fig} but for samples drawn from Group 1 ``observations'' of our simulated satellite debris which is not guaranteed to move with a single velocity within the survey volume.
\label{sim_mcmc.fig}
}
\end{figure}

Figure \ref{sim_mcmc.fig} shows the result of applying the same MCMC analysis to synthetic observations of our simulated data sets for ``Group 1'' as indicated in Figure \ref{vrvsl.fig}  (with number of data points indicated in the middle panel).
These data sets are not guaranteed to be composed of stars distributed with isotropic velocity dispersion about a single mean velocity within the survey volume. 
Nevertheless, the MCMC approach recovers the mean motion of all our synthetic samples within its indicated uncertainties. 


\subsection{Predictions of Other Associated Debris: Connecting to the Clouds}

\label{predict.sec}

Once a mean motion is derived, it is possible through orbit integration to explore whether an observed group might be associated with other phase-space structures elsewhere in the Galaxy.
It might be expected that these predictions could have limited use, given the large uncertainty ($> 100$km/s in Figs. \ref{ideal_mcmc.fig} and \ref{sim_mcmc.fig}) in the motions derived from our small data sets.
This concern is assessed in this section by comparing the known distribution of debris in our simulations of the progenitor of Group 1 with the simple orbit integrations from the derived motions.

Predictions for debris locations were produced by performing integrations for $n_{\rm predict}=10$ orbits at each of the the angular positions of the $n_*=15$ stars in the ``observed'' Group 1 data set (i.e. for a total of $n_{\rm predict}\times n_*$ orbits). 
The full initial conditions for each $n_{\rm predict}$ orbits associated with each star are assigned by: (i) choosing a distance from the Sun at the star's observed angular position by drawing a  random value from a Gaussian distribution whose standard deviation matches the uncertainties \citep[assumed to be 20\%, as estimated in][]{sheffield12}; (ii) choosing a velocity from the PDF defined by the MCMC chains; (iii) projecting the chosen velocity along the line-of-sight; and (iv) deciding whether to adopt the chosen velocity by using an acceptance/rejection technique so that the final $n_{\rm predict}$ line-of-sight motions follow the average and error (assumed to be 10 km/s) estimated for the star; (v) returning to step (ii) if the velocity is rejected or step (i) if it is accepted until all initial conditions for a given star are generated.

The orbits are integrated in the potential in which the simulation was actually run \citep[described in][]{bullock05}, which included smoothly-evolving bulge, disk and halo components. (For the integration, the potential parameters were frozen at the values used in the final time-step of the simulation.)
In reality, of course, we will not know the potential this accurately.
Repetition of the integrations in other potentials \citep[e.g., the triaxial-halo model of][]{law10} produced broadly similar results, suggesting that the current uncertainties in the orbits themselves, not the potential, are the dominant factor limiting the accuracy of the predictions.
However, there were systematic differences in the details of the predictions in the various potentials (e.g., location of apocenters on the sky) that could become interesting once we have more certain estimates of initial conditions.
 
\begin{figure}
\epsscale{0.9}
\plotone{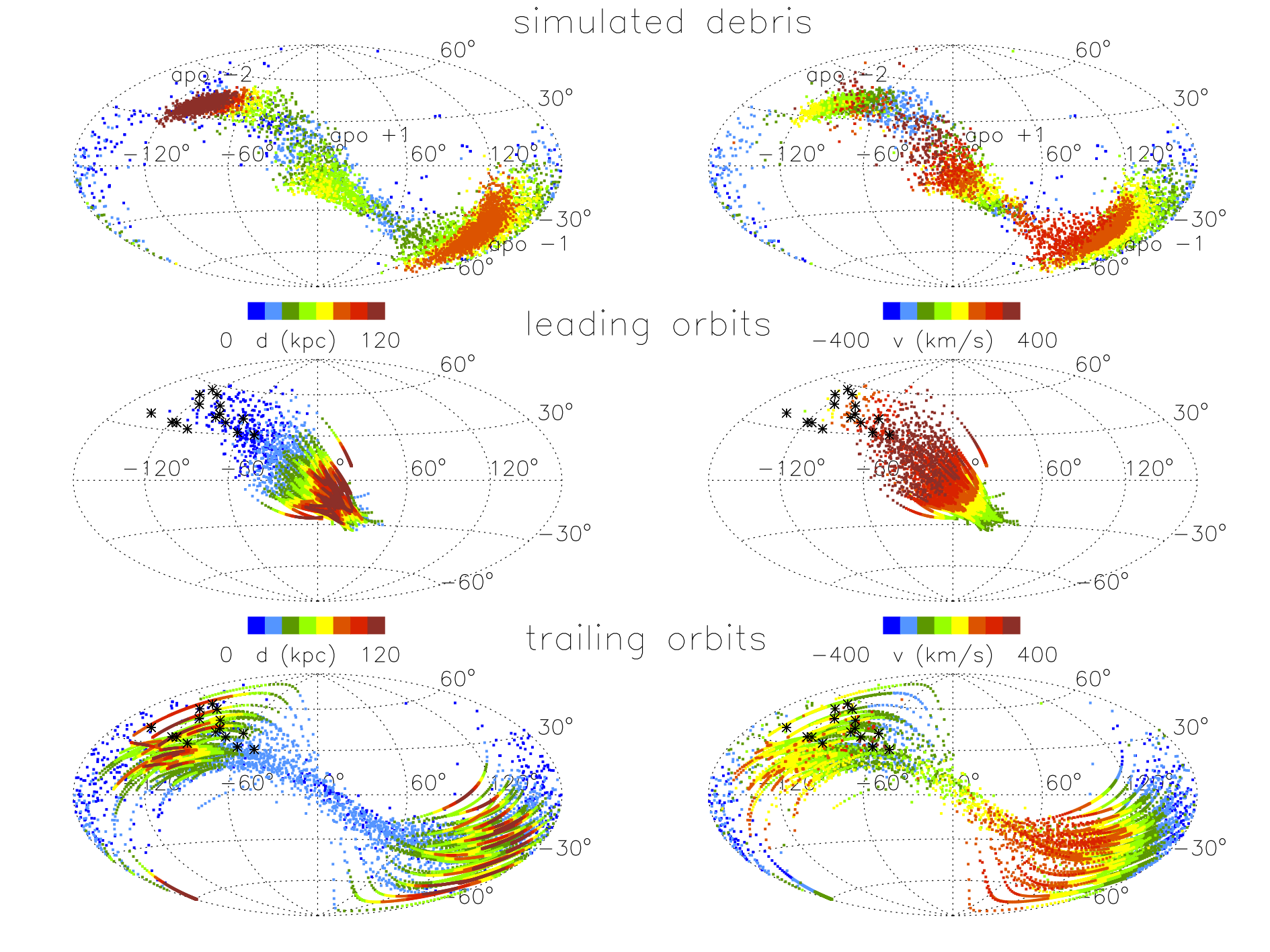}
\caption{Aitoff projections of observed properties of  our simulations (top panels) and for leading/trailing (middle/lower rows) test particle orbits color coded by distance (left hand panels) and line-of-sight speed (right-hand panels). 
For the simulations, a random set of equally-weighted debris particles are plotted.
The black symbols show the position of the particles which are each used to generate 10 test-particle orbits.
The orbits are plotted at equal time intervals far enough into the past and future to encompass the same number of apocenters seen in the simulations.
\label{predict_group1.fig}
}
\end{figure}

Figure \ref{predict_group1.fig} illustrates the results of this process by comparing all-sky projections of the true distance and line-of-sight velocities in our simulated debris for the progenitor satellite of Group 1 (top panels) with those predicted from integrations of the full space motion from the application of the MCMC to our ``observations'' (middle and bottom panels).
The middle panels show the results for integrations starting from the black stars into the future (``leading'' orbits) and
the lower panels are for integrations into the past (``trailing'' orbits). 
The top panels show localized overdensities at large distances (in yellow, orange and red in top left hand panel) corresponding to the apocenters labeled in Figure \ref{sats.fig}.
Note that there is a systematic trend in the distances to the apocenters in this simulated debris that reflects the trend in orbital energy increasing from the leading (apocenter +1 --- green and yellow points) to trailing (apocenters -1 --- orange points --- and -2 --- red points) debris \citep[see][for full explanation]{johnston98}.
Diffuse ``pericentric streams'' connect these apocenters at closer distances (blue points in top left hand panel) where the stars are moving quickly past the Galactic center.
In the top right-hand panels the ``apocentric clouds'' exhibit strong velocity gradients \citep[as also seen in][]{johnston08}.

There are some striking similarities and also striking differences between the top and two lower sets of panels.
The orbital integrations are remarkably successful at outlining the general location of the pericentric streams and apocentric clouds on the sky, the sense of velocity and distance gradients and the magnitude of the velocities.
However, the exact angular positions of the apocenters are not accurately pinpointed (particularly in the trailing debris) and the predictions overall are not as well collimated as the simulations.
Lastly, the distance to the apocenters are systematically too high in the leading orbits because there is no gradient in the orbital properties in our predictions, unlike for real debris from a disruption event.

Overall we conclude that, even with large uncertainties on our derivation of the space-motion of local debris we can make useful predictions for where further members of a debris structure might be found --- in particular for the regions of the sky to search for apocentric clouds.

\section{Results II: Application to Observations}

\label{obs.sec}

The results in the previous section demonstrated that the linear features seen in the $(v_{\rm los}/\cos b)-l$ plane in our simulated surveys correspond to pericentric streams from moderately massive and recent accretion events.
In this section we explore the implications of adopting this interpretation for features found in our observed M giant data set. 
In further work we aim to confirm the reality of these local groups and solidify any preliminary connections made with follow-up observations of both the pericentric streams and apocentric clouds.


\subsection{Summary of Known Structures}

\label{structures.sec}
There are two categories of structures that might be compared to analyses of our data using the tools developed in Sections \ref{motion.sec} and \ref{predict.sec}: nearby phase-space clumps and distant spatial overdensities.

\begin{figure}
\epsscale{0.9}
\plotone{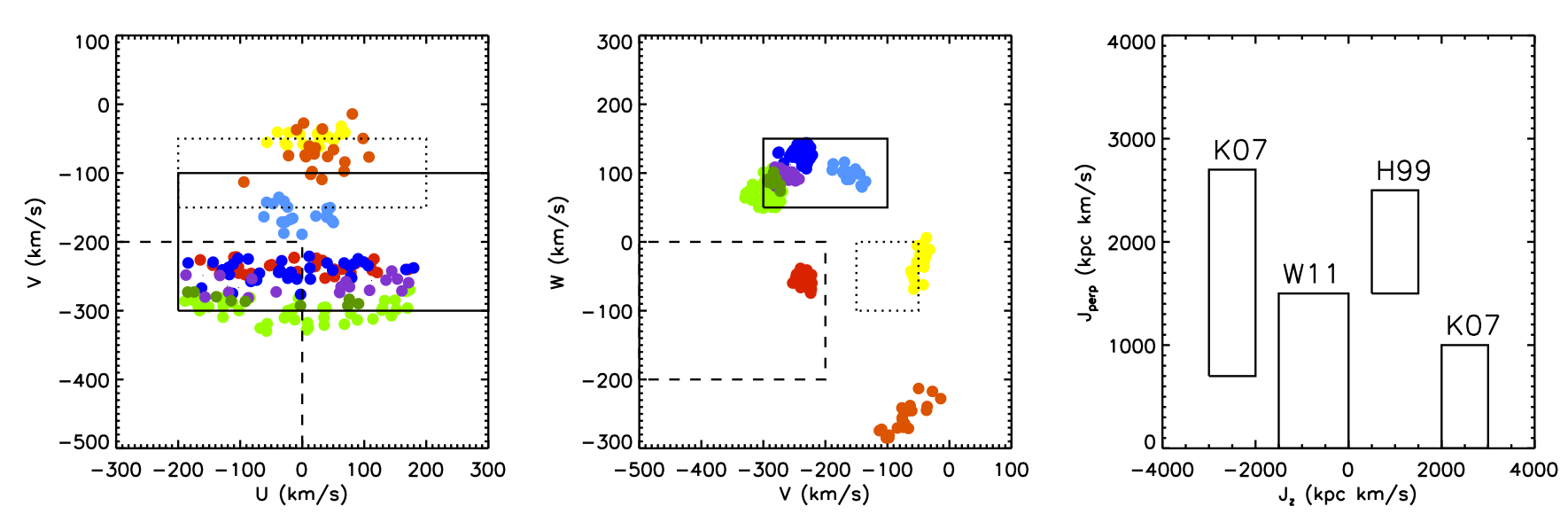}
\caption{The left and middle panels show the $(U,V,W)$ ($=(-v_x,v_y,v_z)$) motions of groupings of stars in the heliocentric frame found in the SDSS survey within $\sim$2.5 kpc of the Sun \citep[colored points, see ---][]{klement09}, as well as the regions of groupings found in motion and metallicity by  \citet{majewski96}. The right panel shows the locations of four distinct groups in angular momentum found by \citet{helmi99}, \citet{kepley07} and \citet{williams11} 
\label{move.fig}
}
\end{figure}

Specifically, once we have derived the full space motion for the postulated pericentric streams as implied by our velocity sequences, we can ask whether there is a match of that stream to any groups found in other surveys that may have full-space motion estimates. 
Figure \ref{move.fig} summarizes the locations of groups whose members have been identified on the basis of similar orbits, either because of their similar motion (left-hand and middle panels) or because of similar angular momentum (right-hand panel).
\citep[As noted above, many of these previous observational studies adopted a co-ordinate system with $U=-v_x$ in the direction of the Galactic anticenter so we present our results in that frame in this and all subsequent plots --- see][]{kepley07}.
Three of the studies are similar in the sense that they all select metal-poor stars within 1-2 kpc of the Sun (i.e. to find a halo sample), but each uses a different tracer: 
\citet{helmi99} (referred to as H99 in Figure \ref{move.fig}) uses a sample of red giant and RR Lyrae stars within 1kpc of the Sun; 
\citet{kepley07} (referred to as K07) uses a combination of red giants, RR Lyares and Red Horizontal Branch stars within 2.5kpc of the Sun; 
and \citet{klement09} uses a sample metal poor main sequence stars selected from SDSS.
The remaining two studies found groups at slightly larger distances (few-10 kpc):
the \citet{majewski96} groups are from a survey to $B\sim 22.5$ at the North Galactic Pole, that reached to distances of $>5$kpc;
and \citet{williams11} (referred to as W11) finds stars in the RAVE data set out to 10 kpc.

\begin{figure}
\plotone{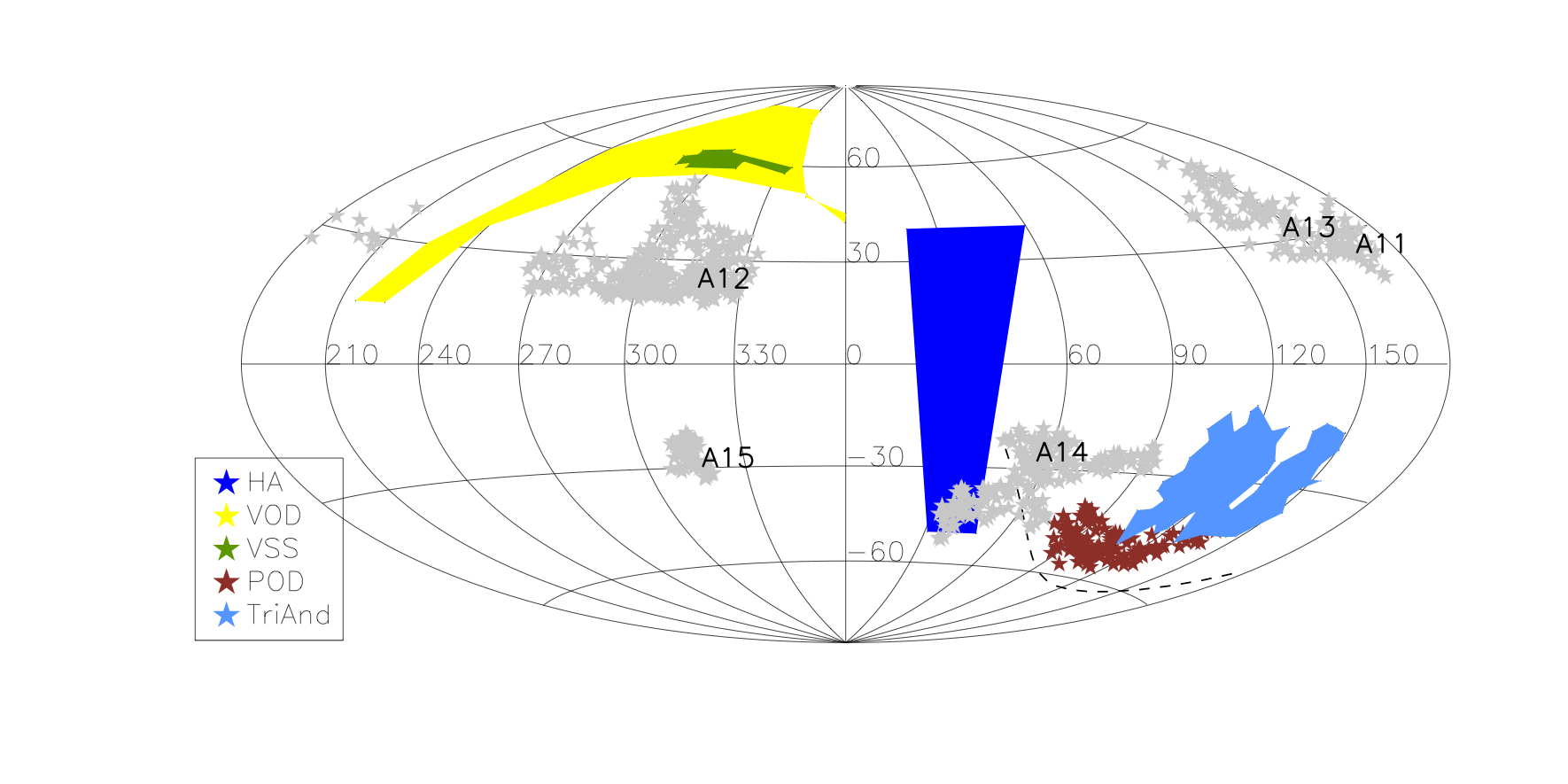}
\caption{Aitoff projection showing sky positions of significant cloud detections (see text for references and discussion of individual features). The gray stars indicate the locations of other overdense groups of M giants found in \citet{sharma10} whose nature has yet to be confirmed. \citep[Adapted from figure 1 of][]{rochapinto10}
\label{clouds.fig}}
\end{figure}

Figure \ref{clouds.fig} shows a summary of the locations of known clouds far beyond these nearby moving groups:
\begin{description}
\item{The Triangulum-Andromeda Cloud (TriAnd)}
was discovered  as an overdensity in the 2MASS M giant catalogue 16-25 kpc away subtending an area of at least $50^\circ \times60^\circ$ in the constellations of Triangulum and Andromeda  \citep[][]{rochapinto04}, and subsequently also detected in main sequence turnoff stars \citep{majewski04}.  \citet{rochapinto04} observed 36 TriAnd M giant candidates with the Bok 2.3-m to derive a metallicity of the structure ([Fe/H]$\sim$ -1.2) and to map its RV structure. The velocity distribution of these stars is reminiscent of models of debris clouds, with a strong gradient in velocity across the cloud going from positive to negative velocities \citep{johnston08}. Chemical tagging through echelle spectroscopy of six M giant members of this cloud confirmed that TriAnd has dSph-like $[\alpha$/Fe] and s-process patterns similar to, though still distinct from, those seen in Sgr and Monoceros stream M giants \citep{chou11} --- a result supporting an origin for TriAnd that is different from either of these two other halo substructures  \citep[in contradiction to the hypothesis of][that the TriAnd Cloud is an apogalaticon piece of the more nearby Mon Stream]{penarrubia05}.

\item{The Virgo Overdensity (VOD)} 
was first mapped extensively by \citet{juric08}, who found a density enhancement in the SDSS stellar catalogue centered at $(l,b)=(300,65)$ and covering more than 1000 deg$^2$.
The cloud crosses the locations of some prior detections of spatial substructure in RR Lyraes and main sequence turnoff stars \citep{vivas01,newberg02}, with which it may be associated.
Stars in SDSS in the VOD are spread out in the range of $6<Z<20$ kpc above the plane with density actually increasing towards the Galactic plane at the inner boundary (6 kpc) of the survey volume.
\citet{juric08} also found a suggestive over-density of M giant stars selected from the 2MASS catalogue coincident with this region.
The SDSS stars in the VOD have intermediate metallicities (-1.6$<$[Fe/H]$<$-1).  
While there has been no systematic kinematic survey covering the face of the clouds, several targeted radial velocity studies  have picked up groups of stars in the direction of Virgo that may be part of the VOD or may be the signature of other streams crossing this region of the sky \citep{duffau06,newberg07,vivas08,prior09}.
For example; in a spectroscopic study of F stars in the direction of Virgo, \citet{newberg07} find peaks at $v_{\rm GSR}$=130 km s$^{-1}$, -76 km s$^{-1}$, and -168 km s$^{-1}$; and
\citet{vivas08} isolate RR Lyraes in the direction of the VOD with three distinct peaks in RV at $v_{\rm GSR}$=215, -49, and -171 km s$^{-1}$.  

\item{The Hercules-Aquila Cloud (HerAq)}
was discovered by \citet{belokurov07} as a stellar overdensity using SDSS photometry and found to stretch from $30^{\circ}<l<60^{\circ}$, with detections both above and below the Galactic plane at distances of 10-20 kpc.
When these authors examined the RV distribution in the region $20^\circ < b < 55^\circ$ and $20^\circ < l < 75^\circ$ they found the peak of the radial velocities to lie at $v_{\rm GSR}$=180 km s$^{-1}$, distinct from those of the Galactic disk or halo. 
Subsequently, \citet{watkins09} looked at the distribution of RR Lyraes in ``Stripe 82'' of SDSS (which was repeatedly observed, and hence allowed the detection of variable stars), and found that almost 60\% of these lay in the stripe at locations where it crossed the HerAq cloud.
These stars had distances in range 5-60 kpc, and metallicities in range [Fe/H]$\sim$ -1 to -2., with an average [Fe/H]=-1.42 $\pm$ 0.24. 

\item{The Pisces Overdensity (POD)}
was initially discovered as a knot of RR-Lyrae stars in SDSS ``Stripe 82'' at a distance of $\sim$90 kpc \citep{sesar07,watkins09}. When \citet{sharma10} applied a specialized group-finder \citep{sharma09} 
to the 2MASS M giant catalogue they identified an amorphous cloud of stars  covering hundreds of square degrees at distances of 80-100 kpc from us and crossing SDSS ``Stripe 82'' 
at a location coincident with the RR Lyrae clump. The M giant cloud plausibly 
provides a much more extensive view of the same structure. Although prior work interpreted 
the RR Lyrae clump as a low-surface-brightness dwarf or star stream \citep{kollmeier09,sesar10},
the M giant cloud points towards POD's origin as a massive, recent accretion event \citep[because its stellar populations include M giant stars -- see][]{sharma11a} that was disrupted along a very eccentric orbit \citep[because of the debris morphology -- see][]{johnston08}.

\end{description}

Figure \ref{clouds.fig} also indicates the locations of five more groups (labeled A11-A15) identified as local overdensities in M giants from the 2MASS catalogue by \citet{sharma10}. The reality of these structures is less certain than the clouds listed above since they do not have confirmed associations with other velocity or metallicity groups. Nevertheless, we include them in order to investigate whether there is a possible association with our own local velocity groups.

In addition, the location of the {\it Virgo Stellar Stream} (VSS) is plotted \citep{newberg02,vivas03,duffau06}; the VSS is a debris structure overlapping the VOD on the sky, but with rather different morphology and tighter velocity dispersion.
\citet{casetti-dinescu09} report on proper motions for stars in a field 7$^\circ$ west of the density peak, and find one star that is consistent in RV and metallicity with the VSS.
Orbit integrations suggest an orbital pericenter of $\sim$11kpc and an apocenter of $\sim$90 kpc.

Other well-known structures are not included in this figure because their orbits are not expected to pass within our survey volume, and hence we do not expect to find nearby counterparts of those distant debris structures in our radial velocity data.
This includes all of the known morphologically extended streams, such as Sagittarius \citep{majewski03}, the Orphan Stream \citep{belokurov06a} and the globular clusters streams \citep{grillmair06}. 

\subsection{Group D}

\label{grpD.sec}

The stars used in our next analyses are highlighted with black symbols in the lower panel of Figure \ref{obs.fig}, with circles indicating the ones that had additional high-resolution spectroscopic data.
The chosen stars were required to: (i) to stand out from the bulk of the disk stars in the survey by having either distinct $V_{\rm GSR}/\cos(b)$ at their longitude $l$ or (in the case of indistinct RV) much lower metallicity  ([Fe/H]$< -0.6$); and (ii) fall on the trend of velocity outlined by other stars in the group.
For Group D, a total of 11 stars in our medium resolution spectroscopic survey fulfilled these combined constraints and 9 of those had high-resolution spectra.

\subsubsection{Derivation of Full Space Motion}
\label{grpD_motion.sec}

\begin{figure}
\plotone{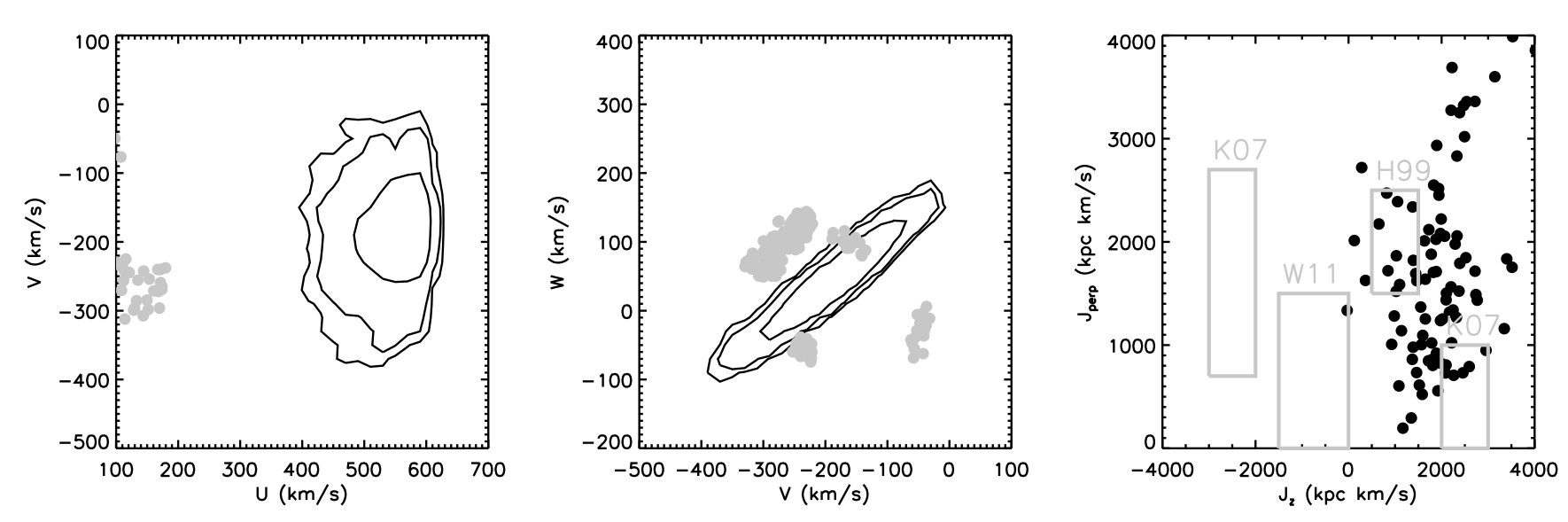}
\caption{Left and middle panels show projections of the PDF for the full space motion of Group D from the MCMC analysis in the $(U,V,W)=(-v_x,v_y,v_z)$ co-ordinate system.  Right panel shows the angular momenta of the 90 particles generated from the PDF for the 9 stars with distances derived from high-resolution spectra. Gray symbols and boxes in all panels ouline the observed groups shown in Figure \ref{move.fig}.
\label{grpD_vel.fig}
}
\end{figure}

Figure \ref{grpD_vel.fig} shows the results of running the MCMC analysis (see section \ref{motion.sec})  on the line-of-sight velocity data for Group D in the heliocentric rest-frame.
The boundary of possible values for $(U,V,W$) returned by the MCMC algorithm was set at the local escape speed \citep[$\sim 600$ km/s, as found in RAVE ---][]{smith07}.
If the boundary on acceptable velocities is dropped, the error contours become even more extended in the $U$-dimension, with a preferred value slightly greater than the escape speed.
Note that the small number of stars means that the uncertainty on the mean motion is very large.

The PDF for $(U,V,W)$ from Figure \ref{grpD_vel.fig} was used to generate initial conditions for our orbit integrations (as outlined in section \ref{predict.sec}). 
The motions were translated to a Galactocentric frame assuming values $(v_{x,\odot},v_{y,\odot},v_{z,\odot})=(11.10,12.24,7.25)$km/s for the Sun's motion relative to the Local Standard of Rest (i.e., of a closed orbit at the Sun's distance form the Galactic Center), and $\Theta_0=236$ km/s for the motion of the LSR itself \citep{bovy09,schonrich10}.
Initial conditions were generated  for $n_{\rm predict}=10$ particles scattered about the mean distance and line-of-sight velocities for each of the the $n_*=$9 stars in Group D (i.e. that had high-resolution spectra, and hence more accurate distance estimates).
The positions in the Galactic frame were calculated assuming a distance of $R_\odot=8$ kpc for the Sun from the Galactic Center.
The right panel of Figure \ref{grpD_vel.fig} shows the angular momentum of these particles with their wide spread giving some idea of the uncertainty in the motion of this group.

A comparison of Figure \ref{grpD_vel.fig} with Figure \ref{move.fig} (the observed group data repeated in gray in Figure \ref{grpD_vel.fig}) demonstrates that there are no clear associations between Group D and any of the prior known groups derived from other nearby velocity surveys. 
This could be due to  several factors: (i) the debris form Group D may be too low in density to have been seen before; (ii) the debris may simply not cross the smaller volume explored by several of these surveys; or (iii) the group identification procedure in other the surveys may focus on stars of too low-metallicity to contain a significant fraction of M giants.
Some of our particles are within the region in angular momentum of the H99 group (right hand panel), but the H99 stars were found to be on orbits with apocenters less than 20kpc, and the orbits for our particles have much larger apocenters (see next section).
Overall, our derived motion is much higher than for the prior known groups, and this motion reinforces that debris associated with Group D reaches far out into the Galaxy, which emphasizes the plausibility of an association with clouds.

\subsubsection{Predictions for and Possible Associations with Clouds}
\label{grpD_predict.sec}

The $n_{\rm predict}\times n_*$ orbits were integrated in  the potential outlined in  \citet{law10}:
\begin{equation}
        \Phi_{\rm disk}=- {GM_{\rm disk} \over
                 \sqrt{R^{2}+(a+\sqrt{z^{2}+b^{2}})^{2}}},
\label{diskeqn}
\end{equation}
\begin{equation}
        \Phi_{\rm sphere}=-{GM_{\rm sphere} \over r+c},
\label{bulgeqn}
\end{equation}
\begin{equation}
        \Phi_{\rm halo}=v_{\rm halo}^2 \ln (C_1 x^2 + C_2 y^2 +C_3 x y + (z/q_z)^2 + r_{\rm halo}^2)
\label{haloeqn}
\end{equation}
where $M_{\rm disk}=1.0 \times 10^{11}$ $M_{\odot}$, $M_{\rm sphere}=3.4 \times 10^{10}$ $M_{\odot}$, $a=6.5$ kpc, $b=0.26$ kpc, $c=0.7$ kpc and $r_{\rm halo} = 12$ kpc. 
The normalization of the dark halo mass via the scale parameter $v_{\rm halo}$ is  specified (for a given choice of $q_1$, $q_2$, and $\phi$) by the requirement that the speed of the Local Standard of Rest be $v_{\rm LSR} = 220$ km/s.
Note that this is a different value for $v_{\rm LSR}$ than used in our transformations, but, given the uncertainties in predicted motion, this difference is not expected to affect our results.
$R/r$ are cylindrical/spherical radii respectively, and the various constants $C_1, C_2, C_3$ are given by
\begin{equation}
C_1 = \left(\frac{\textrm{cos}^2 \phi}{q_{1}^2} + \frac{\textrm{sin}^2 \phi}{q_{2}^2}\right)
\end{equation}
\begin{equation}
C_2 = \left(\frac{\textrm{cos}^2 \phi}{q_{2}^2} + \frac{\textrm{sin}^2 \phi}{q_{1}^2}\right)
\end{equation}
\begin{equation}
C_3 = 2 \textrm{sin}\phi \textrm{cos} \phi \left( \frac{1}{q_{1}^2} - \frac{1}{q_{2}^2}\right)
\end{equation}
This form for the dark halo potential describes an ellipsoid rotated by an angle $\phi$ about the Galactic $Z$ axis, in which $q_1$ and $q_2$ are the axial flattenings along the equatorial axes and $q_z$ is the axial flattening perpendicular to the Galactic disk, and $\phi = 0^{\circ}$ corresponds to $q_1/q_2$ coincident with the Galactic $X/Y$ axes respectively, 
and increases in the direction of positive Galactic longitude.

The calculations were repeated several times in this potential: (i) with a spherical halo (i.e. $q_1=q_2=q_z$=1); (ii) with a triaxial halo ($q_1=1.38$, $q_2=1.0$, $q_z=1.36$ and $\phi=1.69$) where the axis ratios and orientation had been adjusted to best-match the phase-space structure of the debris associated with the Sagittarius dwarf galaxy \citep[as derived by][]{law10}; and (iii) for cases (i) and (ii) but with initial conditions generated from the PDF from MCMC runs where the velocities were allowed to exceed the local escape speed.

\begin{figure}
\plotone{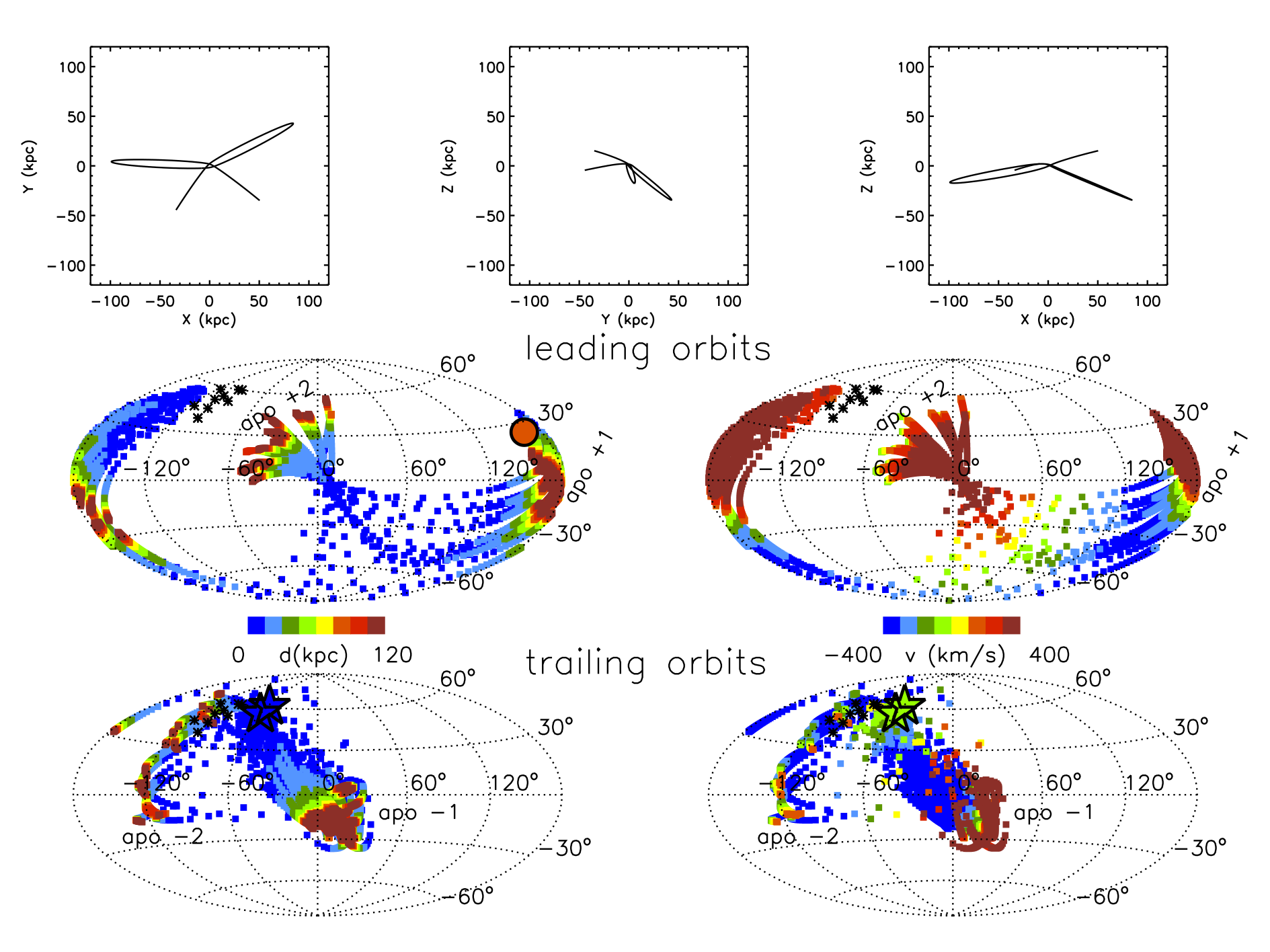}
\caption{Sample orbit (top panels) and aitoff projections (lower panels) for the leading (middle row) and trailing (bottom row) orbits calculated from initial conditions generated from the PDF's shown in Figure \ref{grpD_vel.fig}. The top panels follow the integrations for $\pm$1.5 Gyrs, and the lower panels show debris for two leading/trailing apocenters. Aitoff projections on the left/right are color-coded for distance/velocity.
Black points indicate the positions of stars in Group D.
Colored stars are detections of velocity groups, with the colors showing distance and velocity  estimates from \citet{newberg07} and \citet{vivas08}.
The circle indicates the location and distance of group A11 from \citet{sharma10}.
\label{aitoff_grpD.fig}
}
\end{figure}

Figure \ref{aitoff_grpD.fig} presents the results for the integrations of initial conditions within the Galactic escape speed run in the spherical version of the potential through two (future/past) apocenters leading/trailing the observed stars.
The apocenters in the leading (middle panels) and trailing (lower panels) orbits stand out clearly in red in the left-hand panels at large distances from the Galactic center. 
As found in Section \ref{predict.sec}, although these distances can be substantially overestimated for the leading debris (and possibly underestimated for the trailing debris), we expect the angular locations of the apocenters --- in particular the ones closest in orbital phase to that of the observed stars (indicated by $+1$ and $-1$) --- to provide useful indications of the locations of associated apocentric clouds. 
Indeed the locations of the closest apocenters was similar (i.e.,  to within tens of degrees) for our test cases in both spherical and triaxial potentials.
Moreover, any unbound orbits were predicted to leave the Galaxy in the same portion of the sky as these first apocenters.
In contrast, the morphology and location of the apocenters further apart in orbital phase from the observed data (i.e.,  $+2$ and $-2$) were found to vary more substantially (in some case by up to 100 degrees in angular separation) across our test cases. 
While this calls into question the utility of our current predictions for searching for debris at such large separations in orbital phase, it does emphasize the power that any future associations that can be made across more than a single oscillation in radial phase would have on constraining the potential \citep[as also noted by][]{varghese11}.

It is apparent from a comparison of the middle and lower panels in Figure \ref{aitoff_grpD.fig} with Figure \ref{clouds.fig} that none of the predicted apocenters exactly match the locations of well-known clouds.
The predicted apocenters are at too great distances and low latitudes for most of the surveys.
However, one of the possible groups found by \citet{sharma10} (named ``A11'') is only slightly offset in angular position from the predicted location of the first apocenter along the leading orbits (around $(l,b)=(164^\circ,25^\circ)$), and at a distance slightly less than that predicted (consistent with expectations for leading debris) --- as indicated by the orange circle in middle-left panel of the figure.
Moreover, our predicted apocentric location is actually slightly beyond the edge of \citet{sharma10}'s survey boundary, plausibly explaining the perceived agular offset.
We conclude that, while \citet{sharma10} felt that the proximity of A11 to both the disk and the edge of their chosen survey area called into question its authenticity as a real structure,
our own tentative connection to local debris provides motivation for a follow-up spectroscopic survey. 
If such a survey revealed a strong velocity gradient across the structure going from positive to negative $v_{\rm GSR}$ both the validity of the A11 group, and its connection to our own Group D could be confirmed.

Another plausible connection that can be made is to one of the velocity structures seen in the direction of the VOD and VSS --- though not to those specific overdensities themselves.
The colored stars in the bottom panels indicate the locations, distances and line-of-sight speeds for one  of these peaks derived from spectroscopic surveys in the direction of the VSS \citep{newberg07,vivas08,brink10} that best matches our predictions.
The coincidence between our predictions and these observations is striking.

It is also interesting to note that, while we do not match the radial velocity of the VSS in this region, the orbital properties derived using proper motion measurements for the VSS itself by \citet{casetti-dinescu09} are remarkably similar (apocenter of $\sim 90$ kpc, pericenter of $\sim 11$ kpc and an orbital inclunation of 58$^\circ$) to the one illustrated in the top panels of Figure \ref{aitoff_grpD.fig}.  This agreement is somewhat circumstantial since there is a wide range of orbital properties among the predicted orbits more generally.
However, if we could indeed connect these two separate pericentric passages, we would have an even greater chance of knowing where to look for related clouds.


\subsection{Group E}

\label{grpE.sec}

As for Group D, the stars used in our analyses for Group E were chosen to stand out from the disk in either motion or abundances or both, and to be part of the distinct sequence picked out by eye in the $v_{\rm GSR}/\cos b$ {\it vs} $l$ plane.
They are highlighted with black symbols in the lower panel of Figure \ref{obs.fig}, with circles indicating those particular stars that have additional high-resolution spectroscopy.
A total of 13 stars in our medium resolution spectroscopic survey fulfilled these combined constraints and 5 of those have high-resolution spectra.
In addition, in a subsequent observing run at MDM observatory we found yet another star that fell along the Group E sequence, and we include it in our analysis and all figures.

\subsubsection{Derivation of Full Space Motion}

\label{grpE_motion.sec}

\begin{figure}
\plotone{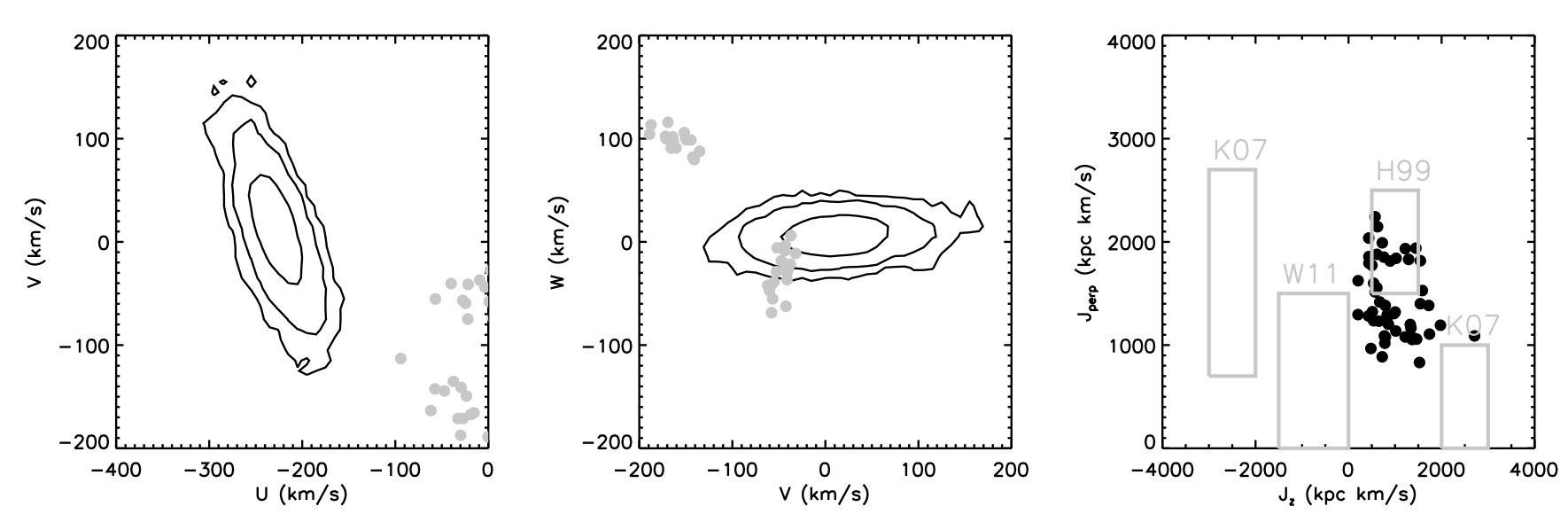}
\caption{As Figure \ref{grpD_vel.fig}, but for Group E stars.
\label{grpE_vel.fig}
}
\end{figure}

The left and middle panels of Figure \ref{grpE_vel.fig} shows the results of running the MCMC analysis (see section \ref{motion.sec})  on the line-of-sight velocity data in the heliocentric rest-frame.
The PDF contours enclose a smaller area of velocity space than Group D, that is far from the local escape speed.
The right panel shows the angular momentum of particles generated from these PDF's.
As with Group D, there is no clear-cut association between any of the local, previously known moving- or angular momentum groups and our derived properties for Group E.

\subsubsection{Predictions for and Possible Associations with Clouds}
\label{grpE_predictions.sec}

\begin{figure}
\plotone{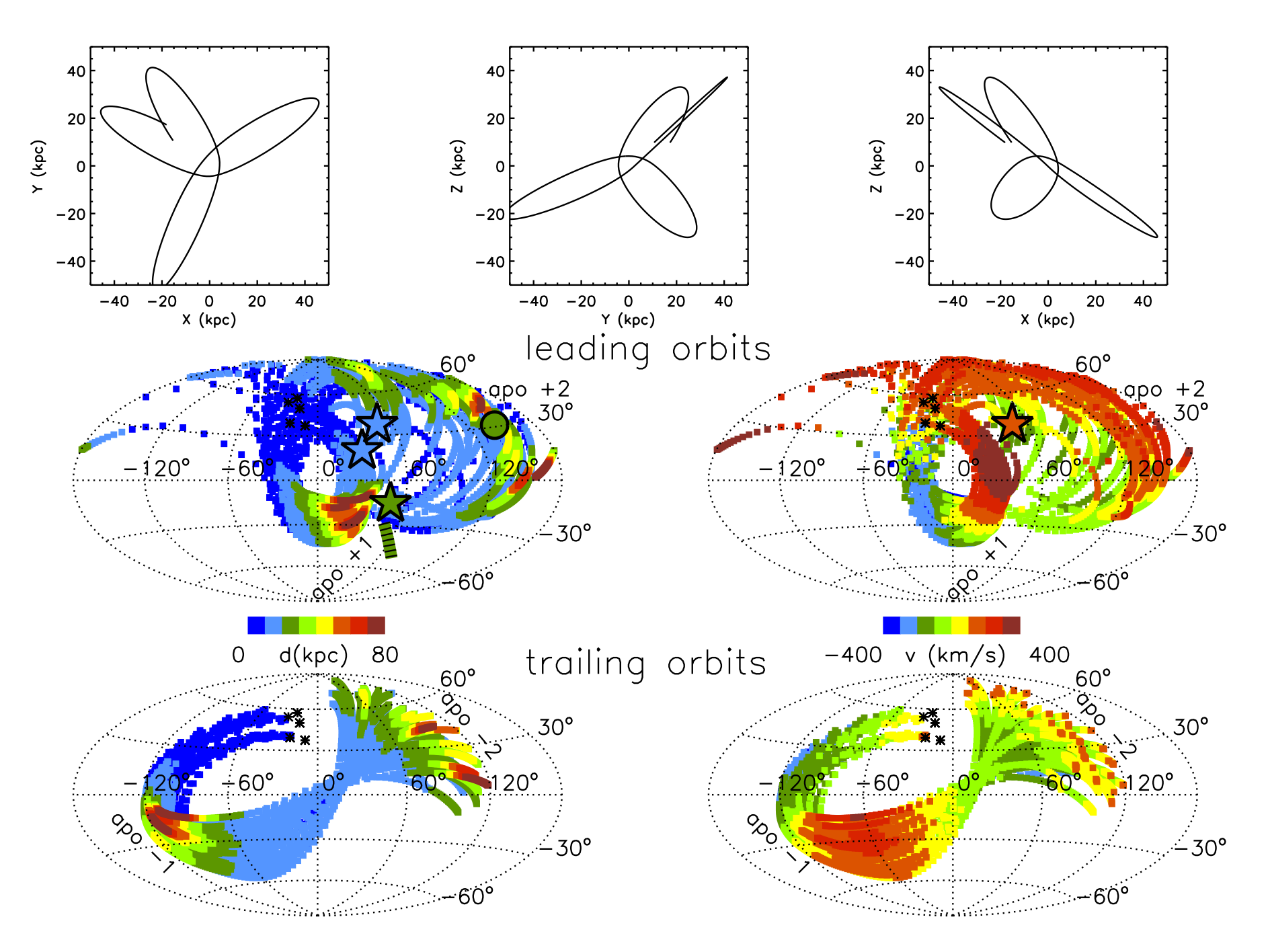}
\caption{Sample orbit  (top panels) and aitoff projections (lower panels) for the leading (middle row) and trailing (bottom row) orbits calculated from initial conditions generated from the PDF's shown in Figure \ref{grpE_vel.fig}. The top panels follow the integrations for $\pm$1.5 Gyrs, and the lower panels show debris for two leading/trailing apocenters. Aitoff projections on the left/right are color-coded for distance/velocity.
Black points indicate the positions of star in Group E.
Colored stars are detections of the HerAq cloud reported in \citet{belokurov07}.
The green line in the middle-left panel indicates the overdense region, and distance of RR Lyraes found in SDSS "Stripe 82" \citep{watkins09}.
The circle shows the location and distance of M giant group A13  reported in \citet{sharma10}.
\label{aitoff_grpE.fig}
}
\end{figure}

Figure \ref{aitoff_grpE.fig} presents the results for the integrations of initial conditions generated from the MCMC PDF's for the motion of Group E in the version of our potential with a spherical halo component.
As for Group D, the predictions up to the first leading/trailing apocenters are very similar in the different potentials, but quite different at orbital phases farther away from the initial conditions of the integrations.

The most promising cloud association for Group E from our orbit predictions is where the leading orbits plunge across the disk plane to  their first apocenters just below the disk around $(l,b)\sim (30^\circ,-20^\circ)$ --- as apparent in the middle left panel of Figure \ref{aitoff_grpE.fig}. The crossing and apocenter overlap the region where the Hercules-Acquila cloud was first discovered using SDSS photometry \citep[see Figure 1 of][for a striking view]{belokurov07} . 
While SDSS does not cover the location of our predicted apocenter, \citet{belokurov07} demonstrated that low-latitude SEGUE observations at a longitude $l=50^\circ$  clearly show the star counts to be denser and more distant below the disk (at $(l,b)\sim (50^\circ,-15^\circ)$) than just above, in the same sense as our predictions.
The colored stars indicate where \citet{belokurov07} detected and made distance estimates for this overdensity, and the colored, dashed line indicates the locations and distances of  RR Lyraes in SDSS Stripe 82 \citep{watkins09}.
The bulk of the population in both studies is found in the range 10-20 kpc, although \citet{watkins09} detected stars as far as 60 kpc.
While these populations are somewhat closer than the apocenters of the orbit integrations, we expect the leading orbits to overestimate the distances.
Moreover, our orbit integrations suggest the densest, central region may sit outside the SDSS or SEGUE footprints, at lower Galactic latitude and longitude.

\citet{belokurov07} also looked at the region $20^\circ < b < 55^\circ$ , $20^\circ < l < 75^\circ$ and found a population of stars associated with this overdensity with a peak line-of-sight motion $\sim 180$ km/s (see orange star in leading orbit's right-hand panel). Once again, the observed region is offset from our stream predictions, but the motion is in the same sense expected.

The distance and velocity observations encircle our leading orbital paths, suggesting that the pericentric stream represented by Group E, as well as the SDSS photometric and spectroscopic detections, might both be skirting the edge of a much larger structure.
If this is indeed the case, then we should be able to find the densest region of the cloud around the predicted apocenter in the  M giant catalogue.
Note that both \citet{belokurov07} and \citet{watkins09} estimated average metallicities for their detections of [Fe/H]$\sim$-1.5 (with a tail to values as high as [Fe/H]=-1) , which is too low for a large fraction of M giants in the stellar population.
This, along with their faint magnitude cut ($K_s>10$), may explain why HerAq was not picked up in the same regions as SDSS by the \citet{sharma10} group-finding analysis of the M giants from 2MASS.
\citet{sharma10} {\it did} find a group of M giants at low, negative Galactic latitude and longitude (labeled ``A4'' in their work), but the majority were clearly part of the Sgr trailing debris stream.

\begin{figure}
\plotone{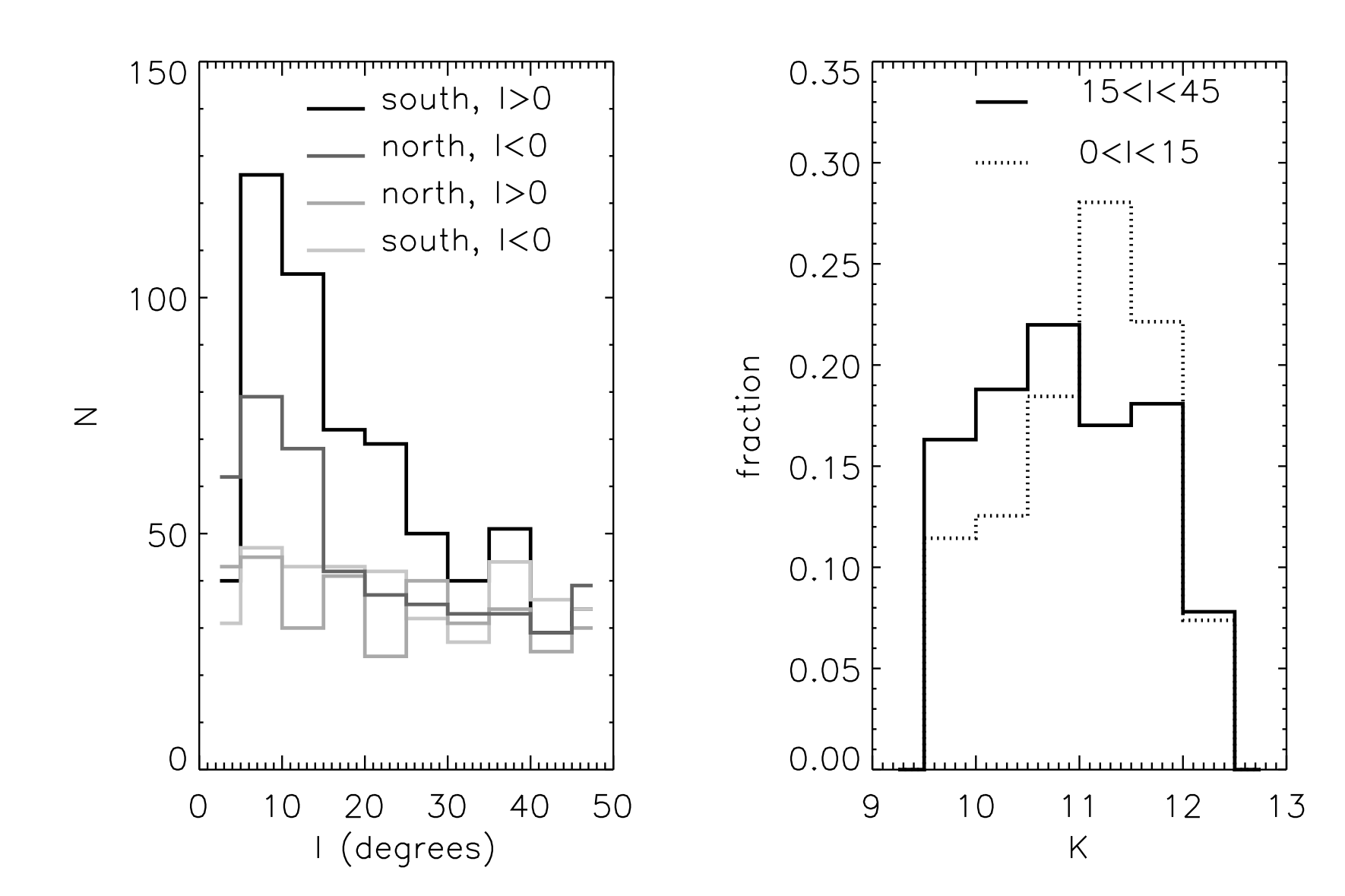}
\caption{The left-hand panel shows M giant star counts in the latitude range $30^\circ < |b| < 50^\circ$ for the northern Galactic hemisphere at negative $l$ (dark grey line) and southern Galactic hemisphere at positive $l$ (solid bold line). lighter grey lines indicate counts in the corresponding north/south positive/negative $l$ regions.
The right hand panels show distribution in apparent $K_s$ magnitudes for M giants in the bold histogram in the left hand panel in the longitude ranges $15^\circ < l<40^\circ$ (solid) and $0^\circ < l < 15^\circ$ (dotted). 
\label{heraq_hist.fig}
}
\end{figure}

Figure \ref{heraq_hist.fig} re-assesses the nature of overdensities in the M giants by plotting the number of stars in the latitude range $30<|b|<50$ as a function of Galactic longitude (left hand panel).
The lines represent star counts in four regions: the northern and southern Galactic hemispheres, at positive and negative longitudes.
These distributions are expected to be similar in the absence of any asymmetries about the Galactic center --- as is the case for $l >0^\circ$ in the north and $l<0^\circ$ in the south (lighter grey lines).
The presence of Sgr debris in the other fields is obvious in the $\sim 10^\circ$-wide peaks at $l\sim -10^\circ$ in the north (dark grey line) and $l\sim 10^\circ$ in the south (black line line). 
In addition, in the south there is an apparent extension of the Sgr overdensity to  higher longitudes, as seen by the counts  around $l\sim 20^\circ$.
To assess whether this extension is indeed part of the Sgr debris, the right panel plots the distribution of $K_s$ apparent magnitudes for the M giants  represented by the solid, black histogram in the left panel in the two ranges $0^\circ<l<15^\circ$ and $15^\circ < l < 40^\circ$.
Comparison of these distributions clearly shows that the stars at higher longitudes (where we expect HerAq to be) are at systematically brighter magnitudes than those in the Sgr region, suggesting that they could indeed be a distinct debris structure.
Moreover, \citet{majewski03} commented on a local peak in star counts at this position along the Sgr trailing debris ($30^\circ < \Lambda_\odot < 50^\Lambda_\odot$ in their Figure 13). In a  subsequent spectroscopic survey, \citet{majewski04} found a dozen stars in this area, within 5 kpc of the Sgr orbital plane, but with velocities inconsistent with the Sgr tails and instead falling in the range $-100 < v_{\rm GSR} < 100$ (see their Figure 2a) --- just as might be expected for HerAq cloud debris turning around at apocenter.

Overall, we conclude that Group E may plausibly represent the pericentric stream associated with the HerAq cloud, and with a potential diffuse distribution of M giants around $(l,b)=(20^\circ,-30^\circ)$ in particular.
A velocity map across the face of the cloud could confirm this association: Sgr debris in the region would have GSR velocities decreasing slowly as $|b|$ increases along the tail; in contrast, our predictions for the Group E debris in this region are for a steep gradient from negative to positive velocities as $l$ increases.


Finally, looking further along the leading orbits to the second apocenter in our predictions we find another connection, to structure ``A13'' discovered in the group-finding analysis of the more distant 2MASS M giant sample by \citet{sharma10}.
The structure coincides in both position and distances with our expectations (filled circle in middle left panel).
Nevertheless, we consider this connection more speculative than the one made to HerAq, both because our predictions are less consistent in different potentials at these wide separations in orbital phase, and because the reality of the A13 structure itself is considered uncertain.
Discovery of a systematic velocity gradient across structure A13, with motions along the line-of-sight decreasing towards lower Galactic latitudes would bolster both the physical reality of A13 and its association with Group E.

\subsection{Alternative Interpretation of Group E}

As noted above, the predictions for locations of other debris associated with both Group D and E rely on the {\it assumption} that the sequences picked out by eye from Figure \ref{obs.fig} result from a pericentric stream passing with a single velocity through the survey volume. 
In a concurrent independent study of nearby K-giants, a sequence at a similar location in longitude/velocity space to Group E has been identified \citep{majewski12}.
However, this sequence has been interpreted, using N-body simulations, as resulting from stars associated with the disruption of the globular cluster $\omega$Cen, turning around near the apocenters of their orbits.
These different scenarios for the K and M giant sequences can be distinguished using observations of proper motions (stars connecting to clouds would have much high proper motions) and 
based on the chemistry of these stars, which show [Ba/Fe] patterns unique to $\omega$Cen 
\citep{smith00,majewski12}.

\section{Summary and Conclusions}
\label{conclusions.sec}

This paper presents a first investigation into the possibility of connecting sequences of stars found in velocity surveys of the nearby Galaxy to 
large scale inhomogeneities in photometric surveys of the outer Galactic halo.
If such connections can be made, they would bolster our confidence in the physical interpretation of these features as pericentric streams associated with apocentric clouds of tidal debris.

Our analysis of simulations of satellite accretion demonstrates that:
\begin{itemize}
\item sequences in velocity across the sky found in spectroscopic surveys may indeed be the signature of streams of debris passing through the local volume;
\item the full space motions of the streams can be estimated directly from these sequences;
\item even with $\sim$15 stars in a sequence, integrations of the (very uncertain) derived full space motions in any reasonable potential provide useful predictions for the locations and motions of other associated debris;
\item the predictions differ significantly as the shape of the potential varies once integrations are separated by more than one radial orbit from the initial conditions.
\end{itemize}

Applying these ideas to two possible sequences (Groups D and E) found in a spectroscopic survey of M giants reveals plausible associations with various prior detections of 
cloud-like overdensities --- in particular, for Group D to structure A11 and, for Group E to HerAq and A13.

We conclude that follow-up spectroscopic surveys of these structures should allow us to confirm or rule out these tentative associations. 
If associations are confirmed, these debris systems would be the first comprehensive maps of satellite disruption around the Milky Way along eccentric orbits, filling in a knowledge of Galactic accretion history that is currently missing.
Moreover, the maps would have comparable extent to those of Sgr's debris streams, and hence provide strong constraints on the Galactic potential.
Finally, knowledge of the velocity of streams of stars passing at high speed through the solar vicinity would provide insight into the interpretation of direct dark matter detection experiments and help constrain the nature of dark matter itself.

\acknowledgements
The work of KVJ and AAS on this project was funded in part by NSF grants AST-0806558 and AST-1107373. AAS acknowledges the  support of Columbia's Science Fellows. All the authors thank the other members of the ``Cloud Hopping'' team for their ongoing contributions to this program: Rachael Beaton, Jeff Carlin, S\'ebastien L\'epine, Ricky Patterson, Ricardo Mu\~noz, Whitney Richardson and Helio Rocha-Pinto.


\end{document}